\documentclass[12pt]{article}

\usepackage[left=2.7cm,bottom=2.7cm,right=2.7cm,top=2.7cm]{geometry}

\usepackage{amsmath,amsfonts}
\usepackage{array}
\usepackage[caption=false,font=normalsize,labelfont=sf,textfont=sf]{subfig}
\usepackage{textcomp}
\usepackage{stfloats}
\usepackage{url}
\usepackage{verbatim}
\usepackage{mathtools}
\usepackage{algorithm}
\usepackage{algorithmic}
\usepackage{threeparttable}
\usepackage{adjustbox}   
\usepackage{rotating}
\usepackage{caption} 
\usepackage{cases}
\usepackage{enumerate}
\usepackage{tabularx}
\usepackage{multirow}
\usepackage{bm}
\usepackage{authblk}
\usepackage{indentfirst}
\usepackage{setspace}
\usepackage{color}
\usepackage[utf8]{inputenc} 
\usepackage[T1]{fontenc}    
\usepackage{hyperref}       
\usepackage{url}            
\usepackage{booktabs}       
\usepackage{amsfonts}       
\usepackage{nicefrac}       
\usepackage{microtype}      
\usepackage{lipsum}		
\usepackage{graphicx}
\usepackage{doi}
\usepackage{amssymb}                  
\usepackage{mathrsfs} 
\usepackage[resetlabels]{multibib}
\usepackage[bottom]{footmisc}
\newtheorem{definition}{Definition}
\newtheorem{lemma}{Lemma}
\newtheorem{proposition}{Proposition}
\newtheorem{theorem}{Theorem}

\newtheorem{proof}{Proof}
\newtheorem{remark}{Remark}
\newtheorem{assumption}{Assumption}

\usepackage[utf8]{inputenc}
\usepackage{natbib}

\usepackage{tikz}
\usetikzlibrary{patterns}
\usetikzlibrary{decorations.pathreplacing,calligraphy}
\usetikzlibrary{patterns.meta}
\usetikzlibrary{arrows.meta}
\tikzdeclarepattern{
  name=mylines,
  parameters={
      \pgfkeysvalueof{/pgf/pattern keys/size},
      \pgfkeysvalueof{/pgf/pattern keys/angle},
      \pgfkeysvalueof{/pgf/pattern keys/line width},
      \pgfkeysvalueof{/pgf/pattern keys/pat},
  },
  bounding box={
    (0,-0.5*\pgfkeysvalueof{/pgf/pattern keys/line width}) and
    (\pgfkeysvalueof{/pgf/pattern keys/size},
0.5*\pgfkeysvalueof{/pgf/pattern keys/line width})},
  tile size={(\pgfkeysvalueof{/pgf/pattern keys/size},
\pgfkeysvalueof{/pgf/pattern keys/size})},
  tile transformation={rotate=\pgfkeysvalueof{/pgf/pattern keys/angle}},
  defaults={
    size/.initial=5pt,
    angle/.initial=45,
    line width/.initial=.4pt,
    pat/.initial=dashed,
  },
  code={
      \draw [line width=\pgfkeysvalueof{/pgf/pattern keys/line width},\pgfkeysvalueof{/pgf/pattern keys/pat}]
        (0,0) -- (\pgfkeysvalueof{/pgf/pattern keys/size},0);
  },
}

\usepackage{graphicx} 
\def\bb{\boldsymbol{\beta}}
\def\sj{\setminus j}

\begin{document}

\title{Learning Joint Graphical Models with Computational Efficiency, Dynamic Regularization, and Adaptation}

\author[1]{Shixiang Liu}
\author[2]{Yanhang Zhang\thanks{Shixiang Liu and Yanhang Zhang contributed equally to this work.}}
\author[3]{Zhifan Li}
\author[4]{Jianxin Yin}

\affil[1]{\footnotesize School of Statistics, Renmin University of China}
\affil[2]{\footnotesize Yau Mathematical Sciences Center, Tsinghua University}
\affil[3]{\footnotesize School of Statistics and Mathematics, Zhongnan University of Economics and Law}
\affil[4]{\footnotesize Center for Applied Statistics and School of Statistics, Renmin University of China}

\date{}
\maketitle \sloppy

\begin{abstract}
	Multi-sourced datasets are common in studies of variable interactions, for example, heterogeneous gene networks, cross-domain recommendation, etc., where each source induces a related but distinct dependency structure. 
Joint learning of multiple graphical models (i.e., multiple precision matrices) has emerged as an important tool in analyzing such data.
Unlike separate learning, joint learning can leverage shared structural patterns across graphs to yield more accurate results. 
In this paper, we present an adaptive method named MIGHT (Multi-task Iterative Graphical Hard Thresholding) to estimate multiple graphs jointly.
We introduce an $\ell_0$-based similarity measure, reformulate the joint model as a series of multi-task learning problems, and solve them using a dynamically regularized algorithm based on iterative hard thresholding.
By leveraging the efficiency of the iterative procedure and the inherent parallelizability of the nodewise framework, the proposed method is computationally efficient and amenable to parallel implementation.
Theoretically, we derive the non-asymptotic error bound for the resulting estimator under joint Gaussian graphical models. 
Furthermore, the proposed algorithm is adaptive to heterogeneous column-wise signal strengths: for nodes with strong signals, our estimator achieves improved error bounds and selection consistency adaptively, and also exhibits asymptotic normality {without an additional debiasing step}.
The performance of our method is illustrated through numerical simulations and real-data analysis on a cancer gene-expression RNA-seq dataset.
\end{abstract}
 

	\section{Introduction}
This paper analyzes the joint learning of multiple Gaussian graphical models.
For $p$ fixed variables, assume that we collect observations drawn from Gaussian distributions across $K$ different classes or datasets, denoted as $X^{(1)} \in \mathbb{R}^{n_1 \times p}, \cdots, X^{(K)} \in \mathbb{R}^{n_K \times p}$. Although their corresponding precision matrices, i.e., the inverse covariance matrices, $\Theta^{(1)}, \cdots, \Theta^{(K)} \in \mathbb{R}^{p \times p}$ may exhibit dataset-specific heterogeneity, they often share many common support structures \citep{Guo11joint}.
This structural commonality motivates a joint estimation of these graphical models, rather than separate estimations.

Joint graphical models have significant practical implications in data analysis.
In gene expression studies, we often need to estimate the gene co-expression network (conditional-dependence structure) for each related tissue.
Since different tissues usually share many biological pathways and conditional dependence, 
their network structures often exhibit substantial commonality while still retaining some tissue-specific edges \citep{Seal2023RCFGL, yang2025AJEM}. 
Additionally, after appropriate pre-processing and variance-stabilizing transformations, gene expression data can often be approximated by Gaussian graphical models \citep{Price15JCGS, Ren22gauss}.
Therefore, compared with estimating each graph separately, joint graph learning can more effectively exploit the shared structure, leading to improved accuracy and stability in graph estimation.
Additionally, transfer learning of graphical models has become a heated topic recently.
In this context, joint estimation, which identifies shared structural patterns, can serve as a pre-processing step for transfer learning to assess the similarity between source studies and the target study \citep{Li23transfer, zhao2024trans}.
These findings necessitate the study of joint graph estimation.

In this paper, we propose an efficient and adaptive method for joint graphical model learning. Through theoretical analysis and practical applications, we demonstrate the advantages of our method in statistical inference.

\subsection{Related works}
\noindent \textbf{Joint graphical model.}\quad Joint graphical model estimation has garnered significant attention over the past decade due to its ability to leverage shared structural patterns across multiple datasets.
Pioneering work like \cite{Guo11joint} introduced hierarchical regularization to decompose and identify the common shared and class-specific edges among these precision matrices.
\cite{Witten13joint} proposed two regularization methods: the Group Graphical Lasso (GGL) and the Fused Graphical Lasso (FGL). 
GGL combines an element-wise sparsity penalty, $\sum_{i \neq j} \sum_{k \in [K]} |\Theta_{ij}^{(k)}|$, with a covariate-wise penalty, $\sum_{i \neq j} \left\{\sum_{k \in [K]} \left(\Theta_{ij}^{(k)}\right)^2 \right\}^{1/2}$, to encourage shared sparsity patterns among different graphical models.
In contrast, FGL combines the same element-wise sparsity penalty with a fused-type penalty, $\sum_{k<k'}\sum_{i \neq j} \left| \Theta_{ij}^{(k)} - \Theta_{ij}^{(k')}\right|$, to encourage shared edge values among different graphical models. 
These two foundational methods have inspired many advanced joint estimation methods.
\cite{Price15JCGS} proposed a ridge penalty plus a ridge fused penalty to maintain the similarity among the multi-graphs.
\cite{Gibberd2017} proposed the group-fused graphical lasso method to estimate the time-evolving and piecewise-constant Gaussian graphical model.
\cite{Bilgrau20fused} extended the work of \cite{Price15JCGS} to the case with prior information.
And \cite{wang22JMLR} proposed a convex optimization method with an $\ell_1$ fused penalty under the general case of graph stationarity.
Additionally, joint estimation methods based on the CLIME estimator (proposed by \cite{Cai11CLIME}) \citep{Liu2015, Qiu15timeseries, Cai2016joint}, and Bayesian frameworks \citep{Li2019bayesian, jalali2023bayesian}, have also been investigated.
For a comprehensive review, we refer the reader to \cite{Tsai2022joint}.

\noindent \textbf{Motivation and inspiration.} \quad
While the aforementioned joint estimation methods capture shared structures, some critical challenges remain: (i) high computational cost, (ii) complex tuning procedures, and (iii) the lack of theoretical guarantees for selection consistency and inference.
To address these issues, we begin by considering a group-structured linear model
\begin{equation}\label{eq:ds}
	\begin{aligned}
		y &= \sum_{j=1}^p X_{G_j}\bb_{G_j} + \epsilon, \\
		&\text{with } \underbrace{\sum_{j=1}^p \mathbf{1}(\bb_{G_j} \neq \mathbf0)\leq s}_{\text{group-wise sparsity}}  \text{ and } \underbrace{\sum_{j=1}^p \|\bb_{G_j}\|_0 \leq s s_0}_{\text{element-wise sparsity}},
	\end{aligned}
\end{equation}
with the covariates partitioned into $p$ groups $\{ G_j\}_{j=1}^p$. Within this joint sparse framework, \cite{Simon13SGL} introduced the sparse group lasso, which simultaneously applies both an element-wise penalty $\|\bb \|_1 = \sum_{j=1}^p \|\bb_{G_j}\|_1$ and a group-wise penalty $\sum_{j=1}^p \|\bb_{G_j}\|_2$.
The penalty terms of the sparse group lasso and the GGL exhibit similar patterns, reflecting the analogous multiple sparsity structures in their underlying statistical models.
Recently, there have been many cutting-edge works on the joint sparse model \eqref{eq:ds}: \cite{Tony22SGL} provided theoretical guarantees for parameter estimation of the sparse group lasso method. 
\cite{zhang2023minimax} developed a double sparse iterative hard thresholding procedure to achieve minimax adaptivity.
\cite{abra24ejs} proposed a classifier based on logistic regression that leverages such a joint structure \eqref{eq:ds}.
\cite{Tugnait2021TSP, Tugnait2024TSP} extended the sparse-group-lasso-type penalty to learning multi-attribute graphical models and matrix-valued graphical models, respectively.
Motivated by these developments, we introduce an analytical framework that reformulates the joint graphical model into a series of linear models via the nodewise regression approach \citep{Meinshausen06nodelasso, Jankova2017honest}.

Furthermore, many joint learning studies rely on empirical risk minimization \citep{Witten13joint, Ma16joint, Zhang2024Multi}, and typically analyze statistical properties of the global optimum of an empirical objective.
In contrast, iteration-based analysis does not necessitate reaching a global or local optimum, but rather focuses on the trade-off between iterative optimization error and statistical error \citep{ndaoud2019contributions, she2021Bregman}.
Consequently, this work adopts a dynamically regularized iterative procedure to achieve both computational efficiency and estimation accuracy.

\subsection{Our contributions}

The main contributions of this work are threefold:
\begin{itemize}
	\item
	We characterize the joint model using $\ell_0$-based local sparsity and similarity, which provides a flexible interpretation of the ``similar yet distinct'' principle based on the average frequency of neighbor occurrences.
	We implement this via a method named MIGHT (Multi-task Iterative Graphical Hard Thresholding), which performs joint nodewise learning using a proxy two-step hard thresholding operator.
	This operator intuitively achieves a trade-off between structural consistency and signal heterogeneity.

	\item The MIGHT method achieves high statistical accuracy with computational efficiency. It establishes sharper joint error bounds compared to existing methods.
	We also prove that MIGHT adaptively achieves better estimation accuracy and selection consistency under proper signal conditions.
	Additionally, we demonstrate the asymptotic properties of MIGHT, a key feature for statistical inference.

	\item We implement the MIGHT method in an open-source R package \texttt{ADSIHT}, available at \url{https://cran.r-project.org/web/packages/ADSIHT/index.html}. 
	The comprehensive numerical experiments on synthetic and real datasets demonstrate that our method achieves a more competitive empirical performance than the state-of-the-art methods.
	
\end{itemize}

\subsection{Organization and Notations}

The current paper is organized as follows.
Section \ref{sec2} introduces our method and algorithm. 
Section \ref{sec: theoretical} derives the theoretical guarantees of our method. 
Sections \ref{sec: simu} and \ref{sec: real} demonstrate the advantages of our method through simulations and empirical data, respectively.
Section \ref{sec: conclusion} concludes this paper.
All technical proofs and additional simulation results are provided in the appendices.

Throughout, for the given sequences $a_n$ and $b_n$, we say that $a_n = O(b_n)$ when $a_n \le Cb_n$ for some constant $C>0$, while $a_n = o(b_n)$ corresponds to ${a_n}/{b_n} \rightarrow 0$ as $n \to \infty$.
We write $a_n \asymp b_n$ if $a_n = O(b_n)$ and $b_n  = O(a_n)$.
Let $[m]$ denote the set $\{1,2,\cdots,m\}$, and $\mathbf{1}(\cdot)$ as the indicator function.
For a set $S$, let $|S|$ denote its cardinality.
Let $\mathbf{0}_p$ denote the $p$-dimensional zero vector.
For a vector $\bb\in \mathbb R^p$, we denote by $\|\bb \|_2$ its Euclidean norm, by $\text{supp}(\bb) = \left\{j \in [p] : \beta_j \ne 0 \right\}$ its support index set, and by $\|\bb \|_0 = |\text{supp}(\bb)|$ the number of its nonzero entries. 
For every set $S \subset [p]$, denote by $\bb_S=(\beta_j, j \in S) \in \mathbb{R}^{|S|}$ the subvector of $\beta$ indexed by $S$.  
For a matrix $A$, we denote by $\| A\|_F$ its Frobenius norm, and by $ \Lambda_{\max}(A)$ and $\Lambda_{\min}(A)$ its largest and smallest eigenvalues.
We denote by $A_{ij}$ the element in the $i$-th row and $j$-th column of matrix $A$, and by $A_{\cdot j}$ the $j$-th column of $A$. 
Let $\mathbf I_p \in \mathbb R^{p \times p}$ denote the identity matrix.
We use $C, C_0, C_1,\cdots$ to denote absolute constants, whose actual values vary from time to time.
In the asymptotic analysis, we assume $n, p, K \to \infty$, simultaneously.

\section{Methods}\label{sec2} 

This section introduces the model setup and transforms the joint graphical model into a series of multi-task learning problems. 
A dynamic-regularized iterative hard thresholding-type algorithm is then introduced to solve these problems. 

\subsection{Model setup}\label{subsec: model setup}

\begin{figure*}[!t]
	\centering
	\subfloat[]{\includegraphics[width=3in]{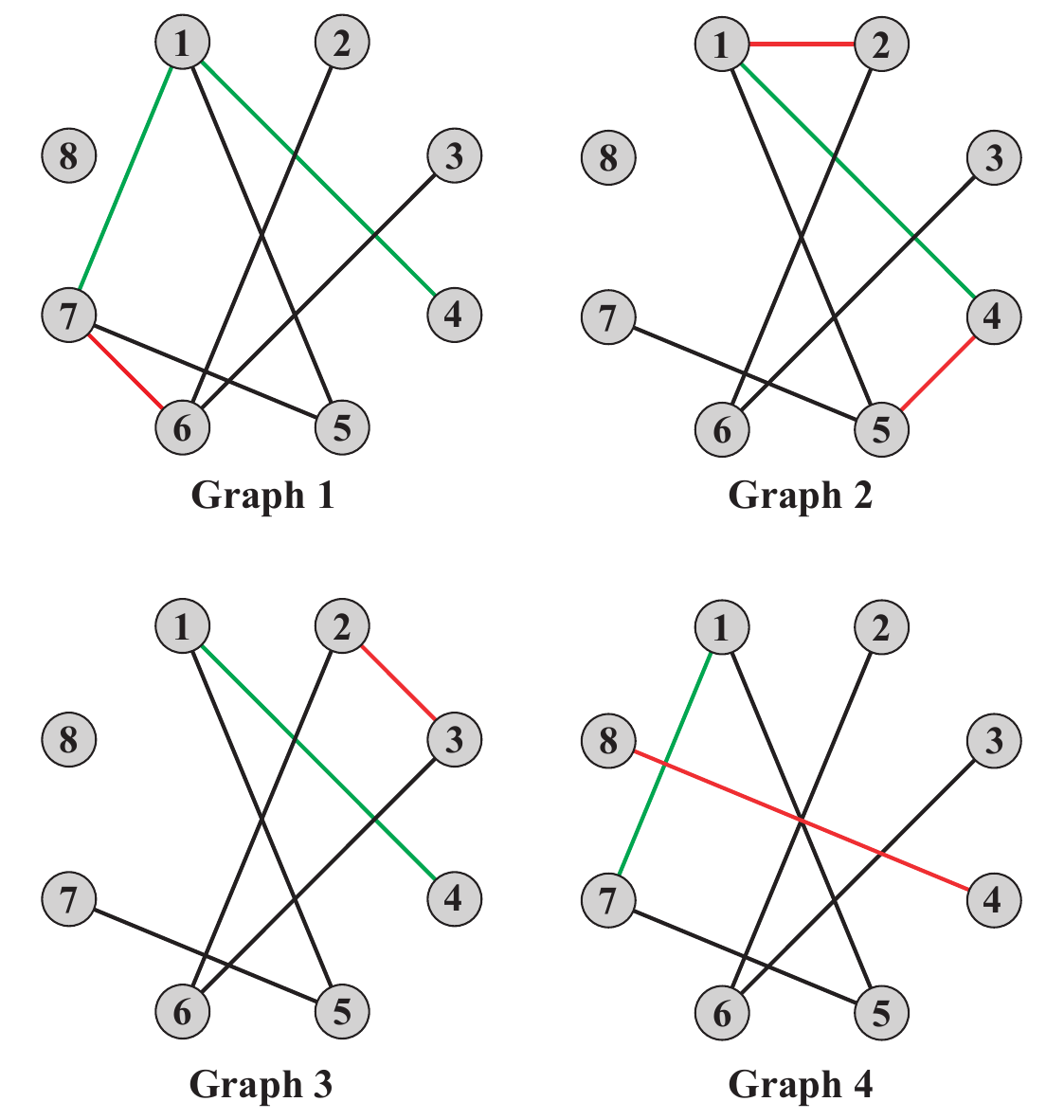}}
	\hfil
	\subfloat[]{\includegraphics[width=3in]{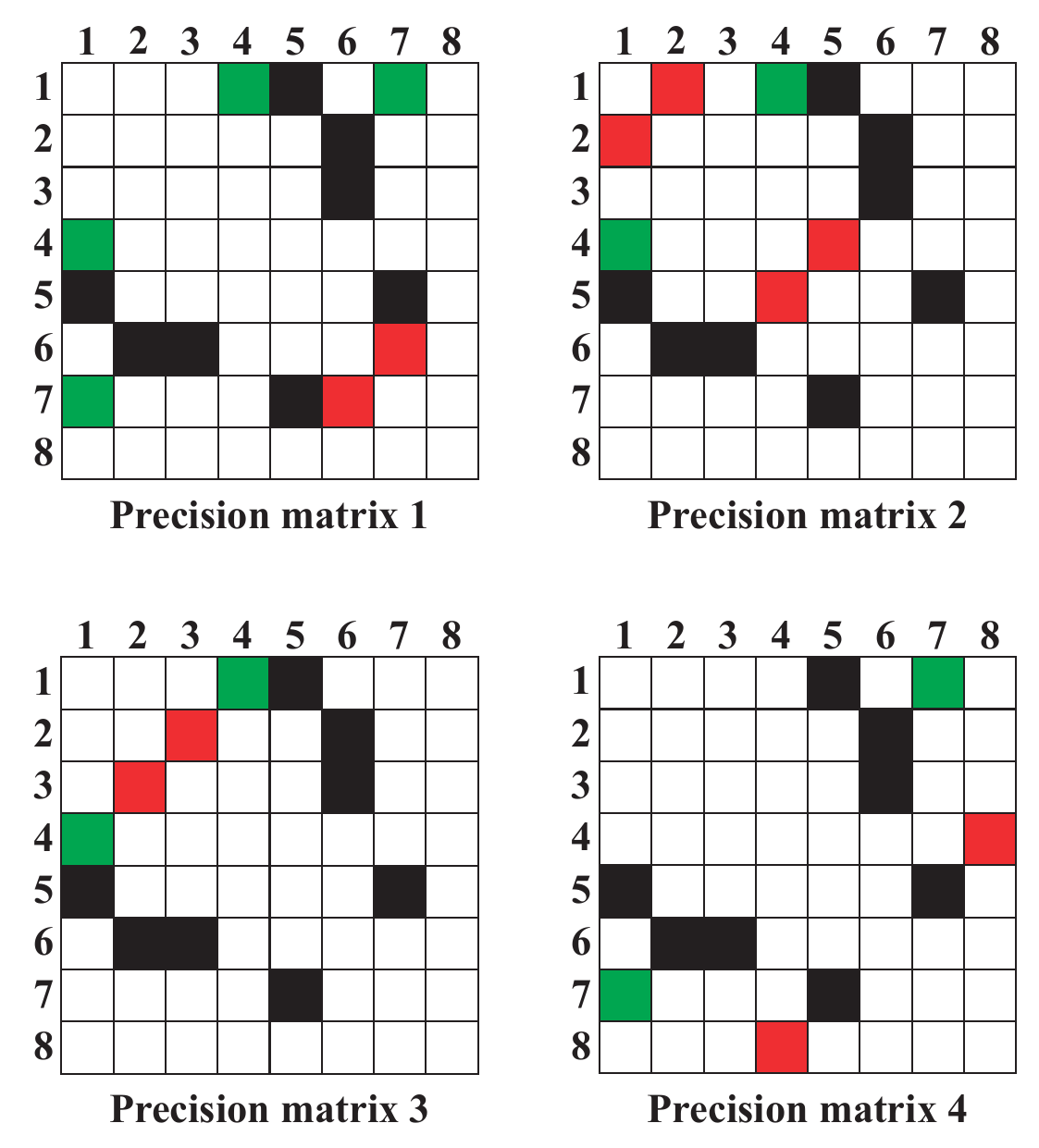}}
	\caption{The joint graphical models with $K=4$ graphs and $p=8$ nodes.
		(a) Edges in these graphs can be divided into three categories: edges universally shared across all graphs (black lines), edges partially shared among subsets (green lines), and edges unique to individual graphs (red lines). 
		(b) Corresponding adjacency matrices for (a).
		Take node 1 as an example, its neighbor sets across the four graphs are $S_1^{(1)} = \{4,5,7\},  S_1^{(2)} = \{2,4,5\}, S_1^{(3)} = \{4,5\}$, and $S_1^{(4)} = \{5,7\}$. The total neighborhood set $S_1= \bigcup_{k=1}^4 S_1^{(k)} = \{2,4,5,7\}$ includes all nodes that connect to node 1 in at least one graph.
		On average, each edge that is connected to node $1$ appears in $ \frac{\sum_{k=1}^4 |S_1^{(k)}|}{|S_1|} = 2.5$ graphs, reflecting the degree of (local) similarity among these graphs. 
	}\label{fig1}
\end{figure*}

Consider the joint estimation of the graphical models across $K$ independent datasets.
For the $k$-th dataset, observations are organized into an $n_k \times p$ matrix $X^{(k)} = \left(x_1^{(k)}, \cdots , x_{n_k}^{(k)} \right)^\top$, where each vector $x_\ell^{(k)} \in \mathbb R^{p\times 1}$ follows an independent $p$-dimensional Gaussian distribution $N(\boldsymbol{\mu}^{(k)}, \Sigma^{(k)})$, for every $\ell \in [n_k]$.
Without loss of generality, we assume $\boldsymbol{\mu}^{(k)}=\mathbf 0_p$.
Denote by $\Theta^{(k)} := \left( \Sigma^{(k)} \right)^{-1} \in \mathbb R^{p\times p}$ the underlying precision matrix of the $k$-th dataset.
For $i \neq j$, the entry $\Theta_{i,j}^{(k)}$ quantifies the conditional dependence between nodes $i$ and $j$ in the $k$-th graph, with $\Theta_{i,j}^{(k)} = 0$ indicating conditional independence of $i$ and $j$ given the remaining $p-2$ covariates \citep{lauritzen1996graphical}.
We denote by $S_j^{(k)}$ the neighbor set of node $j$ in the $k$-th graphical model, i.e.,
$$
S_j^{(k)}: = \left\{ i \in [p]\setminus \{j\}: \Theta_{i,j}^{(k)} \ne 0 \right\},
$$
and let $S_j:= \bigcup_{k=1}^K S_j^{(k)}$ denote the total neighbor set of node $j$ across all $K$ graphs, which includes nodes that connect to node $j$ in at least one graph.
Define $N := \sum_{k \in [K]} n_k$.
We assume that these $K$ graphical models share some common structures, allowing joint estimation to achieve higher accuracy and lower computational cost than separate estimation \citep{Guo11joint, Witten13joint, Tsai2022joint}.
As illustrated in Figure \ref{fig1}, the joint graphs exhibit two structural properties: local edge-sparsity and local neighbor-similarity, which are formalized in the following.

\begin{definition}[Local edge-sparsity]\label{def: sparse}
	We say the joint graphical model $\left( \Theta^{(1)}, \cdots, \Theta^{(K)}\right)$ is locally edge-sparse with parameters $\{s_{j} \}_{j \in [p]}$ if 
	$$ |S_j| = s_j,  \quad \text{for every } j \in [p].
	$$
\end{definition}
We quantify the \textit{local edge-sparsity} by constraining the total neighbor set for each node $j$ respectively (e.g., $s_{1}=4$ in Figure \ref{fig1}). It also provides a sparse neighbor set in each graphical model (since $|S_j^{(k)}| \le |S_j|$ for each $(j,k) \in [p] \times [K]$).

\begin{definition}[Local neighbor-similarity]\label{def: similar}
	We say the joint graphical model $\left( \Theta^{(1)}, \cdots, \Theta^{(K)}\right)$ is locally neighbor-similar with parameters $\{s_{0,j} \}_{j \in [p]}$ if 
	$$
	\frac{\sum_{k=1}^K |S_j^{(k)}|}{|S_j|} = s_{0,j}, \quad \text{for every } j \in [p].
	$$
\end{definition}
We quantify the \textit{local neighbor-similarity} by constraining the average frequency of each neighbor for every node $j \in [p]$.
An $s_{0,j}$ approaching $K$ allows the near-complete neighbor overlap across all graphs, indicating locally homogeneous structure for node $j$.
In contrast, a smaller $s_{0,j}$ encourages each graph to have some individual neighbors of node $j$, indicating greater local heterogeneity among these graphs.

Assume that the joint graphical model exhibits both local edge-sparsity (with $\{s_j\}_{j \in [p]}$) and local neighbor-similarity (with $\{s_{0,j}\}_{j \in [p]}$), which we denote as the joint $(s_j, s_{0,j})_{j\in [p]}$ structure.
We next estimate the precision matrices in a column-by-column fashion, i.e., estimating all the $j$-th columns $\left(\Theta_{\cdot, j}^{(1)}, \cdots,  \Theta_{\cdot, j}^{(K)} \right)$ at once. 
This is inspired by the nodewise regression \citep{Meinshausen06nodelasso}, which reformulates the estimation of a column in a precision matrix as a sparse linear regression problem \citep{Jankova2017honest}. 
For each $(j,k) \in [p] \times [K]$, let $X_j^{(k)}\in \mathbb R^{n_k \times 1}$ denote the observation vector corresponding to the $j$-th covariate, and let $X_{\sj}^{(k)} \in \mathbb R^{n_k \times (p-1)}$ denote the submatrix of $X^{(k)} \in \mathbb R^{n_k \times p}$ excluding the $j$-th covariate.
The next proposition indicates that the column-wise estimation is equivalent to a multi-task learning model.

\begin{proposition}\label{prop: multi-task}
	The joint estimation of all the j-th column $\left(\Theta_{\cdot, j}^{(1)}, \cdots,  \Theta_{\cdot, j}^{(K)} \right)$ is equivalent to the multi-task learning model:
	\begin{equation}\label{eq: multi-task}
		\begin{aligned}
			\underbrace{\begin{pmatrix}
					X_j^{(1)} \\\vdots \\X_j^{(K)}
			\end{pmatrix}}_{=: X_j \in \mathbb R^{N}}
			=& 
			\underbrace{\begin{pmatrix}
					X_{\setminus j}^{(1)} &  \cdots  &\mathbf O_{n_1\times (p-1)}\\
					\vdots&  \ddots& \vdots\\
					\mathbf O_{n_K\times (p-1)}&\cdots&X_{\setminus j}^{(K)}
			\end{pmatrix}}_{=: X_{\setminus j} \in \mathbb R^{N \times (p-1)K}}
			\underbrace{\begin{pmatrix}
					\boldsymbol{\alpha}_{\setminus j}^{(1)}\\ \vdots \\ \boldsymbol{\alpha}_{\setminus j}^{(K)}
			\end{pmatrix}}_{=: \boldsymbol{\alpha}_{\setminus j} \in \mathbb R^{(p-1)K }}
		+
			\underbrace{\begin{pmatrix}
					\epsilon_j^{(1)}\\ \vdots \\\epsilon_j^{(K)}
			\end{pmatrix}}_{=: \epsilon_j \in \mathbb R^{N}},
		\end{aligned}
	\end{equation}
	where $N = \sum_{k \in [K]} n_k$, and 
	\begin{equation}\label{eq: relationship}
		\begin{aligned}
			\boldsymbol{\alpha}_{\sj}^{(k)} &= -\left(\Theta_{j, j}^{(k)}\right)^{-1} \Theta_{\setminus j, j}^{(k)} \in \mathbb R^{p-1}
			,\\
			\epsilon_j^{(k)} &= X_{j}^{(k)} - X_{\setminus j}^{(k)}  \boldsymbol{\alpha}_{\sj}^{(k)} \sim N \left( \mathbf 0_{n_k}, \left(\Theta_{j, j}^{(k)}\right)^{-1} \cdot \mathbf I_{n_k} \right) .
		\end{aligned}
	\end{equation} 
\end{proposition}

Since $S_j^{(k)} =\text{supp}\left( \Theta_{\setminus j,j}^{(k)}\right) = \text{supp}\left(\boldsymbol{\alpha}_{\sj}^{(k)}\right) $, after the column-wise transformation \eqref{eq: multi-task}, the joint $(s_j, s_{0,j} )_{j\in [p]}$ structure is encoded to the coefficient vectors $\left\{\boldsymbol{\alpha}_{\sj} \right\}_{j\in[p]}$.
Additionally, we apply a (sample-based) column-wise scaling in model \eqref{eq: multi-task}, ensuring that each column of $X_{\setminus j}$ has an $L_2$-norm of order $\sqrt{N}$. 
This scaling is used to realize the column normalization condition, which is commonly imposed in high-dimensional regression analysis \citep{Raskutti2011minimax, bellec2018meet}.
The scaled model is then expressed as  
\begin{equation}\label{eq: scaled}
	X_j = Z_{\sj} \bb_{\sj} + \epsilon_j \in \mathbb R^N, \text{ for each } j \in [p],
\end{equation} 
where $\bb_{\sj}\in \mathbb R^{(p-1)K}$ is composed of $\bb_{\sj}^{(1)}, \cdots, \bb_{\sj}^{(K)} \in \mathbb R^{p-1}$, see Algorithm \ref{NodeIHT} and Appendix \ref{Th1.1} for specific rescale operations.
Therefore, our main objective is to estimate the precision matrices $\left( \Theta^{(1)}, \cdots, \Theta^{(K)}\right)$ in a column-by-column fashion, leveraging both local neighbor-similarity and edge-sparsity to achieve accurate and computationally efficient joint estimation.

\subsection{The MIGHT method}

This subsection introduces a nodewise Multi-task Iterative Graphical Hard Thresholding (MIGHT) to solve the scaled model \eqref{eq: scaled}. 
The key to this method is a two-step thresholding operator $\mathcal T_{\lambda, s_0}:\mathbb R^{(p-1)K} \to \mathbb R^{(p-1)K}$ inspired by \cite{zhang2023minimax}, which can be constructed from the following definitions.
\begin{enumerate}
	\item 
	For a vector $\bb = \left( \left(\bb^{(1)}\right)^\top, \cdots, \left(\bb^{(K)}\right)^\top \right)^\top\in\mathbb R^{(p-1)K}$, where each subvector $\bb^{(k)} \in \mathbb R^{p-1}$ corresponds to the $k$-th dataset, we define the \textit{edge-wise} thresholding operator $\mathcal T_{\lambda}^{(1)}: \mathbb R^{(p-1)K} \to \mathbb R^{(p-1)K}$ as:
	$$
	\left\{ \mathcal T_\lambda^{(1)}(\bb)\right\}_{i}^{(k)} = \bb_i^{(k)} \cdot \mathbf 1 \left(|\bb_i^{(k)}| \ge \lambda\right) , 
	$$
	for each $(i,k) \in [p-1] \times [K]$, where $\bb_i^{(k)}$ represents the $i$-th entry of $\bb^{(k)}$.
	
	\item 
	Define the \textit{neighbor-wise} thresholding operator $\mathcal T_{\lambda,s_0}^{(2)}: \mathbb R^{(p-1)K} \to \mathbb R^{(p-1)K}$ as:
	$$
	\left\{ \mathcal T_{\lambda,s_0}^{(2)}(\bb)\right\}^{(k)}_i
	= \bb^{(k)}_i \cdot \mathbf1 \left\{ \sum_{k' \in [K]} \left( \bb^{(k')}_i \right)^2 \ge s_0 \lambda^2  \right\} ,
	$$
	for each $(i,k) \in [p-1] \times [K]$.
\end{enumerate}
The two-step hard thresholding operator can be described as $\mathcal T_{\lambda,s_0}:= \mathcal T_{\lambda,s_0}^{(2)} \circ \mathcal T_{\lambda}^{(1)}$, i.e., $\mathcal T_{\lambda,s_0}(\bb):= \mathcal T_{\lambda,s_0}^{(2)} \left(  \mathcal T_{\lambda}^{(1)}(\bb) \right)$. 
Specifically, in the multi-task learning problem \eqref{eq: scaled} with respect to node $j$, for each potential neighbor $i \in [p]\setminus\{j\}$, if 
$
\sum_{k\in[K]} \left\{ \left( \hat \bb_{i}^{(k)}\right)^2 \cdot \mathbf 1\left(|\hat\bb_{i}^{(k)}| \ge \lambda \right) \right\} \ge s_0\lambda^2
$ holds, 
node $i$ is selected as a neighbor of node $j$, and its {edge} takes the thresholded value
$$
\left\{ \mathcal T_{\lambda,s_0}( \hat\bb ) \right\}_{i}^{(k)} = \hat\bb_{i}^{(k)} \cdot \mathbf 1\left(|\hat\bb_{i}^{(k)}| \ge \lambda \right), \text{ for each }k \in [K].
$$ 
This thresholding mechanism simultaneously controls the edge-sparsity and neighbor-similarity across multiple graphs.

\begin{remark}[Realizing the operator]
	The above two-step operator enforces similarity in a signal-strength-based surrogate sense: A potential neighbor of node $j$ is retained if it exhibits moderately strong edges (with $j$) across sufficiently many graphs, or if it possesses exceptionally strong signals in only a few graphs.
	For example, consider the neighbor relationship $1-2$ in Figure \ref{fig1}, which appears only in Graph 2. 
	Only when the signal $\Theta_{1,2}^{(2)}$ (and its corresponding regression coefficient $\bb_{2,\setminus 1}^{(2)}$) is sufficiently strong can such a task-specific edge be retained.
	Therefore, the two-step thresholding rule does not impose exactly $\ell_0$-based support overlap across graphs.
	Rather, it implements a signal-adaptive notion of similarity: weak task-specific edges are suppressed, while sufficiently strong task-specific edges may be retained because they carry meaningful, albeit heterogeneous, information. 
	Additionally, this operator can be seamlessly integrated with the gradient descent step, ensuring computational efficiency. In contrast, canonical $\ell_0$-based optimal subset selection is NP-hard \citep{natarajan1995sparse}, and therefore is computationally intractable in practice.
\end{remark}

We propose Algorithm \ref{IHT} to solve the scaled model \eqref{eq: scaled}.
Starting from the least-squares loss $\mathcal L (\bb_{\sj}) = \frac2N \left\|  X_j - Z_{\sj} \bb_{\sj}\right\|_2^2$, in each iteration round the algorithm performs a gradient-descent update and then applies the operator $\mathcal T_{\lambda_t, s_0}$ to enforce the desired joint structure.
It consists of two stages: the dynamic thresholding iteration and the fixed thresholding iteration. 
In the dynamic stage (lines 3-7), the thresholding parameter $\lambda_t$ decreases geometrically from an initial value $\lambda_0$ to a pre-specified fixed value $\lambda_\infty$, which effectively balances Type I and Type II errors and yields an initial estimation of $\bb_{\sj}$. Here, $\kappa \in (0,1)$ is a fixed decreasing rate, typically chosen as 0.9 in practice, and the explicit form of constants $C_1^{\text{alg}}, C_2^{\text{alg}}$ will be given in Theorem \ref{th: convergence}.
In the fixed stage (lines 10-13), we construct an appropriate thresholding parameter $\lambda_{\text{fix}}$, which remains fixed in subsequent iterations. This stage refines the initial estimation, ensuring selection consistency and asymptotic properties.

\begin{remark}[Realizing the optimization]
	The gradient step is constructed from a least-squares loss of a joint nodewise regression.
	Therefore, one can interpret Algorithm \ref{IHT} as being related to the following constrained optimization problem (based on the result in Theorem~\ref{th: convergence}):
	\begin{equation*}
		\begin{aligned}
			\text{minimize}_{\boldsymbol\beta_{\setminus j}}: \quad & \frac2N \left\|  X_j - Z_{\sj} \bb_{\sj}\right\|_2^2\\
			\text{subject to}: \quad & \left| \bigcup_{k \in [K]}\text{supp}(\bb_{\sj}^{(k)}) \right| \le C s_j, \\
			& \sum_{k \in [K]}\left| \text{supp}(\bb_{\sj}^{(k)}) \right| \le C s_j s_{0,j}, 
		\end{aligned}
	\end{equation*}
	where \(C\) is a constant depending only on constants $\underline C , \overline C$ (shown in Assumption \ref{assume: eigen}).
	This formulation reflects two joint structural constraints, 
	which are consistent with the local edge-sparsity and local neighbor-similarity as introduced in Definitions \ref{def: sparse} and \ref{def: similar}.
	Notably, the output of Algorithm 1 may not be a global/local optimum of the above problem.
	Nevertheless, we can still provide theoretical guarantees for its output by balancing the iterative optimization error and intrinsic statistical complexity.
\end{remark}

\begin{algorithm}
	\caption{Nodewise Procedure}\label{IHT} 
\begin{algorithmic}[1]
	\REQUIRE $Z_{\sj},~ X_j,~\kappa,~s_{0,j}$
	\STATE Initialize $t=0,~\lambda_\infty =  C_1^{\text{alg}}\sqrt{  \frac{\log K+  (1/s_{0,j})\log p }{N} }$, $\hat \bb_{\sj}^0 = \mathbf 0_{(p-1)K}$  and  $N =\sum_{k \in [K]} n_k$
	\STATE Initialize $\lambda_{0} = C_2^{\text{alg}}$ 
	\WHILE {$\lambda_{t} \geq \lambda_\infty,~$}  
	\STATE $\hat \bb_{\sj}^{t+1} = \mathcal{T}_{\lambda_{t}, s_{0,j} }\left(\hat \bb_{\sj}^{t} + \frac1N Z_{\sj}^\top (X_j - Z_{\sj} \hat \bb_{\sj}^{t} )\right)$ 
	\STATE $\lambda_{t+1} = \kappa \lambda_{t}$
	\STATE $t = t+1$
	\ENDWHILE
	\STATE $\hat s_j = \max\left\{ 1,~\left| \bigcup_{k\in [K]} \text{supp} \left( \left( \hat \bb_{\sj}^{t}\right)^{(k)} \right)\right| \right\}$  \STATE $
	\lambda_{\text{fix}} =  C_1^{\text{alg}} \sqrt{  \frac{\log(\hat s_j K) +  (1/s_{0,j})\log p }{N} }$
	\WHILE {$t \le \left \lceil \frac{\log( \lambda_0/\lambda_\infty)}{\log(1/\kappa)} \right \rceil  + C_3 \log N,~$}
	\STATE $\hat \bb_{\sj}^{t+1} = \mathcal{T}_{\lambda_{\text{fix}}, s_{0,j} }\left(\hat \bb_{\sj}^{t} + \frac1N Z_{\sj}^\top (X_j - Z_{\sj} \hat \bb_{\sj}^{t} )\right)$  
	\STATE $t = t+1$
	\ENDWHILE
	\ENSURE $\hat \bb_{\sj}^t$
\end{algorithmic}
\end{algorithm}

Building on Algorithm \ref{IHT}, we outline the nodewise Multi-task Iterative Graphical Hard Thresholding (MIGHT) Algorithm \ref{NodeIHT} for the joint graphical estimation.
For each node $j$, Algorithm \ref{NodeIHT} firstly constructs a column-rescaled multi-task model \eqref{eq: scaled} (line 4), which leads each column of the design matrix to have an Euclidean norm of order $\sqrt{N}$. 
We then apply Algorithm \ref{IHT} to obtain a nodewise estimator, and finally recover the joint model through the relationship \eqref{eq: relationship}. 
Notably, Algorithm \ref{NodeIHT} allows parallel computation across covariates (lines $6-12$), ensuring computational efficiency.

\begin{algorithm}
\caption{Multi-task Iterative Graphical Hard Thresholding (MIGHT)}\label{NodeIHT} 
\begin{algorithmic}[1]
	\REQUIRE $X^{(1)}, \cdots, X^{(K)},~\kappa$
	\FOR{$k = 1$ \TO $K,~$}  
	\STATE $\hat \Sigma^{(k)} = \frac{1}{n_k} \left( X^{(k)} \right)^\top X^{(k)}$ 
	\STATE $\hat \Gamma^{(k)} = \text{diag}\left( \{\hat \Sigma^{(k)}_{ii} \}_{i \in [p]} \right)$
	\STATE $Z^{(k)} = \sqrt{\frac{C_0 N}{n_k}} X^{(k)}  \left(\hat \Gamma^{(k)}\right)^{-1/2}$ \hfill //Column scaling
	\ENDFOR
	\FOR{$j = 1$ \TO $p,~$}  
	\STATE $\hat \bb_{\sj} = \mathtt{Algorithm1}(Z_{\sj}, X_j, \kappa, \hat s_{0,j})$, with $\hat s_{0,j}$ tuned by \eqref{eq: verzelen} \hfill // Multi-task model
	\FOR{$k = 1$ \TO $K,~$}  
	\STATE $\hat \Theta_{j,j}^{(k)} = n_k \cdot \left\| X_j^{(k)} - Z_{\sj}^{(k)} \hat \bb_{\sj}^{(k)}\right\|_2^{-2}$
	\STATE $\hat \Theta_{\sj, j}^{(k)} = - \sqrt{\frac{C_0 N}{n_k}} \hat \Theta_{j,j}^{(k)} \left( \hat \Gamma_{\sj,\sj}^{(k)}\right)^{-1/2} \hat \bb_{\sj}^{(k)}$ 
	\ENDFOR
	\ENDFOR
	\ENSURE $\hat \Theta^{(1)}, \cdots, \hat \Theta^{(K)}$
\end{algorithmic}
\end{algorithm}

\subsection{Miscellaneous setups}
To ensure adaptability, we employ a variant of the Birg\'{e}-Massart criterion, as introduced by \cite{zhang2023minimax}, to determine the neighbor-similarity degree $s_{0,j}$ in Algorithm \ref{IHT}. 
Concretely, for each $j \in [p]$, let $\hat \bb_{\sj}(s_{0,j})$ denote the estimation obtained under parameter $s_{0,j}$ in Algorithm \ref{IHT}.
We define $\hat s(s_{0,j}) := \max\left\{ 1,~\left| \bigcup_{k\in [K]} \text{supp} \left( \hat \bb_{\sj}^{(k)}(s_{0,j}) \right)\right| \right\}$ as the estimated number of neighbors to node $j$, and $\hat A(s_{0,j}) := \left\| \hat\bb_{\sj}(s_{0,j}) \right\|_0$ as the total number of selected edges (to node $j$) among all $K$ graphs. We select $\hat s_{0,j}$ by the following information criterion:
{\small
\begin{equation}\label{eq: verzelen}
	\begin{aligned}
		\hat s_{0,j} =& \arg\min_{s_{0,j}} \left\{ \log \left( \frac1N \left\|X_j-Z_{\sj}\hat\bb_{\sj}(s_{0,j}) \right\|_2^2 \right)\right. \\
		& \quad + \left.\frac CN \left[\hat s(s_{0,j})\cdot \log p + \hat A(s_{0,j}) \cdot \log \left( K \cdot \hat s(s_{0,j}) \right)\right ] \right\}.
	\end{aligned}
\end{equation}
}

Considering the symmetry of the precision matrix, we adopt the minimum symmetrization approach \citep{Cai11CLIME, Cai2016joint, shu2024nodewise}:
{\small
$$
\hat \Theta_{i,j}^{(k),\text{Sym}} = 
\hat \Theta_{i,j}^{(k)} \cdot \mathbf1\left(\left| \hat \Theta_{i,j}^{(k)} \right| \leq \left| \hat \Theta_{j,i}^{(k)} \right| \right) 
+ \hat \Theta_{j,i}^{(k)} \cdot \mathbf1\left( \left| \hat \Theta_{j,i}^{(k)} \right| < \left| \hat \Theta_{i,j}^{(k)} \right| \right) ,
$$
}
where $\hat \Theta^{(1)}, \cdots, \hat \Theta^{(K)}$ are the estimated precision matrices obtained from Algorithm \ref{NodeIHT}.

\section{Theoretical analysis}\label{sec: theoretical}
This section provides the theoretical guarantee for our MIGHT method under the Gaussian setting.
For simplicity, we assume that the observations of all datasets are of the same order $n$, i.e., $n_1 \asymp \cdots \asymp n_k \asymp n$, and $N = \sum_{k=1}^K n_k \asymp nK$. 
We make the following assumptions:

\begin{assumption}[Gaussianity]\label{assume: gauss}
For each $k \in [K]$, we assume that each row of $X^{(k)}$ is independently drawn from $N(\mathbf0_p, \Sigma^{(k)})$.
\end{assumption}

\begin{assumption}[Bounded eigenvalues]\label{assume: eigen}
There exist two positive constants $\overline C> \underline C >0$ such that
$
\underline C \le \min_{k\in [K]} \Lambda_{\min} \left( \Sigma^{(k)}\right)
\le \max_{k\in [K]} \Lambda_{\max} \left( \Sigma^{(k)}\right) \le \overline C
$.
\end{assumption}

\begin{assumption}[Joint structure]\label{assume: sparse}
Assume the joint graphical model $\left( \Theta^{(1)}, \cdots, \Theta^{(K)}\right)$ has the joint $( s_j, s_{0,j} )_{j\in [p]}$ structure, i.e., has both $\{s_j\}_{j \in [p]}$ local edge-sparsity and $\{s_{0,j} \}_{j \in [p]}$ local neighbor-similarity, with $1\le s_{0,j} \le K,~1\le s_j \le p-1$ for each $j \in [p]$.
\end{assumption}

Assumption \ref{assume: gauss} requires that the observations follow Gaussian distributions. 
Assumption \ref{assume: eigen} excludes singular or nearly singular precision matrices and ensures that the variance in each nodewise regression, i.e., $\text{Var}(\epsilon_j^{(k)}) = 1/ \Theta_{j,j}^{(k)}$, neither diverges nor approaches zero. This assumption is common in studies on graphical models \citep{Ren15ggm, Ma16joint}.
Moreover, Assumption \ref{assume: eigen} together with the sample size condition (specified in Theorem \ref{th: convergence}) ensures that our method still attains desired theoretical guarantees after the sample-based column scaling.
Assumption \ref{assume: sparse} characterizes the node-specific structure of the joint graphs, as we introduced in Definition \ref{def: sparse} and \ref{def: similar}, which indicates $|S_j| = s_j$ and $\sum_{k \in [K]}|S_j^{(k)}| = s_j s_{0,j}$ for every $j \in [p]$.

Furthermore, we emphasize that the theoretical results established in this section remain valid under a broader class of sub-Gaussian designs for each $X^{(k)}$. 
However, for the general sub-Gaussian designs, each precision matrix $\Theta^{(k)}$ should be interpreted as characterizing an inter-regression-based dependence structure, rather than the exact Gaussian conditional dependence.
Therefore, the sub-Gaussian extension should be viewed as an extension of the joint estimation theory, rather than as a general interpretation of the conditional-independence graphs.
Detailed proof and discussion are provided in Appendix \ref{app: sub Gaussian}.

\subsection{Estimation error bound}
The following theorem provides column-wise bounds on the estimation error. For ease of display, for every $j \in [p]$ and $k \in [K]$, we define $\hat S_j^{(k)} := \text{supp} \left( \hat \Theta_{\sj,j}^{(k)}\right)$ as the estimated neighbor set of node $j$ in the $k$-th graph.

\begin{theorem}[Convergence rate and joint structure]\label{th: convergence}
Assume Assumptions \ref{assume: gauss}-\ref{assume: sparse} hold and $n \gtrsim\max_{j \in [p]} \left( s_j\log p + s_j s_{0,j} \log K \right)$.
We apply the MIGHT Algorithm \ref{NodeIHT} with taking
\begin{equation*}
	C_1^{\text{alg}}\ge \frac{4\sqrt{3\overline C} \left( \kappa/\delta+\sqrt3-1 \right)}{ \kappa/\delta -1 },
	\quad C_2^{\text{alg}} \ge \sqrt{ \frac{\overline C^2 (\overline C^2 + \underline C^2 ) }{\underline C^3} },
\end{equation*}
where we define $\delta := \frac{\overline C^2}{\overline C^2 + \underline C^2} \in (0.5,1)$ and assume $1>\kappa > \delta $.
Then, with probability at least $1- C_3 e^{-C_4 \log (pK)}$, we simultaneously have the following properties for every $j \in [p]$:
\begin{enumerate}
	\item Non-asymptotic estimation error bound:
	{\small
		\begin{equation}\label{eq: l2error}
			\begin{aligned}
				\sum_{k=1}^K \frac{n_k}N \left\|\hat \Theta^{(k)}_{\cdot ,j} - \Theta^{(k)}_{\cdot, j} \right\|_2^2 \lesssim &  \frac{ K+ s_j \log p + s_j s_{0,j} \log(s_j K) }{N} . \\
			\end{aligned}
		\end{equation}
	}
	\item Sparse neighbor selection and edge selection: 
	$$
	\left| \bigcup_{k \in [K]} \hat S_j^{(k)} \right| = O(s_j),
	\text{   and   }
	\sum_{k\in [K]}\left| \hat S_j^{(k)} \right| = O(s_j s_{0,j}). 
	$$
\end{enumerate}
\end{theorem}
The joint estimation bound in \eqref{eq: l2error} consists of three components: (i) estimating the $K$ non-zero diagonal entries $\hat{\Theta}_{j,j}^{(k)}$, with rate $K/N$; (ii) estimating $s_j$ neighbors from $p-1$ nodes, with rate $(s_j \log p)/N$; and (iii) estimating $s_j s_{0,j}$ edges from $s_j K$ potential edges, with rate $(s_j s_{0,j} \log(s_j K))/N$.
This decomposition demonstrates that our MIGHT method can adapt to the local $(s_j, s_{0,j})$ joint structure of each node $j$, yielding an adaptive error bound while ensuring the estimator maintains joint sparsity.
Additionally, by combining the node-specific bound \eqref{eq: l2error}, we can obtain the joint estimation bound in the Frobenius norm as
{\small
\begin{equation}\label{eq: frob}
	\sum_{k=1}^K \frac{n_k}N \left\|\hat \Theta^{(k)} - \Theta^{(k)} \right\|_F \lesssim  \sqrt{ \sum_{j \in [p]} \frac { K + s_j \log p + s_j s_{0,j} \log(s_jK) }{N}}.
	\end{equation} }
	\noindent This result offers a more refined error bound than the separation estimation rate $\sqrt{K \cdot \sum_{j \in [p]} \frac{ s_j \log p}{N} }$.
	It is also sharper than the bound in existing joint estimation methods \citep{Cai2016joint, wang17joint}.
	Furthermore, even when prior support structure information is known, the upper bound derived by \cite{Ma16joint}, $\sqrt{\frac{\sum_{j \in [p]} s_j \log p}{N}} + \sqrt{\frac{K \sum_{j \in [p]}s_j }{N}}$, could be less tight than ours in \eqref{eq: frob} in the case when $\sum_{j\in[p]} s_j > \sum_{j\in[p]} s_j  \frac{s_{0,j} \log(s_j K)}{K}$. 
	This case can be readily satisfied when the inter-graph local similarity is relatively low for some of the nodes. 
	This demonstrates the superior accuracy and adaptability of our method.

	\subsection{Support recovery and asymptotic distribution}
	
	We further analyze the performance of our MIGHT estimator in exact support recovery (i.e., neighbor and edge selection consistency) and discuss its asymptotic properties.
	We begin by presenting the minimum signal conditions required for these results.
	
	\begin{assumption}[Minimum signal conditions for node $j$]\label{assume: betamin}
For a node $j \in [p]$, there exists a sufficiently large constant $C_1 >0$ such that
\begin{align}
	\min_{ k \in [K]}  \min_{i\in S_j^{(k)}} \left|\Theta^{(k)}_{i,j} \right|^2
	\ge&  \frac{C_1}{n_k} \left( \frac{\log p}{s_{0,j}} + \log(s_jK) \right), \label{eq: entry}\\
	\min_{i\in S_j } \sum_{k\in [K]} \frac{n_k}{N}
	\left(\Theta^{(k)}_{i,j} \right)^2
	\ge& C_1 \left(\frac{ \log p + s_{0,j} \log(s_j K) }{N} \right). \label{eq: group}
\end{align}
\end{assumption}
For a node $j \in [p]$, this assumption requires: (i) each nonzero per-graph edge $\Theta^{(k)}_{i,j}$ to exceed the per-graph noise level \eqref{eq: entry}, and (ii) each neighbor $i\in S_j$ to have sufficiently large aggregated squared strength across graphs \eqref{eq: group}.
They play complementary roles in guaranteeing exact edge- and neighbor-level support recovery.
These two signal conditions are not implied by one another, because $s_{0,j}$ captures only the average occurrence for the neighbors across graphs, not the exact number of graphs in which a specific $i-j$ edge appears.
Based on this assumption, Theorem \ref{th: supportrecovery} establishes the node-specific selection consistency.

\begin{theorem}[Selection consistency and sharp error bound]\label{th: supportrecovery}
Assume all conditions in Theorem \ref{th: convergence} hold, and define the strong signal node set as $\mathcal A:=\{j\in [p]: \text{ node j satisfies Assumption \ref{assume: betamin} } \}$.
Assume that there exists a constant $\gamma \in (0,1)$ such that $\min_{j \in \mathcal A} \frac{\log p}{s_{0,j}} + \log(s_j K) \ge \gamma \log |\mathcal A|$.
Then, with probability at least $1- C_1e^{-C_{2,\gamma} \min_{j \in \mathcal A}\left( \frac{\log p}{s_{0,j}} + \log(s_jK) \right)}$, for every $j \in \mathcal A$ we simultaneously have the following properties:
\begin{enumerate}
	\item Selection consistency: 
	$$
	\left(\hat S_j^{(1)}, \cdots, \hat S_j^{(K)} \right) = \left(S_j^{(1)}, \cdots, S_j^{(K)} \right).$$
	
	\item Sharper estimation error bound: $$
	\sum_{k \in [K]} \frac{n_k}{N} \left\|\hat \Theta^{(k)}_{\cdot, j} - \Theta^{(k)}_{\cdot, j} \right\|_2^2 \lesssim \frac{ K + s_j s_{0,j}+ \log p}{N}.$$
\end{enumerate}
\end{theorem}

Theorem \ref{th: supportrecovery} shows that our method is inherently signal-adaptive: it adaptively achieves support recovery and sharper estimation rates for those nodes with strong signals, and maintains the results in Theorem \ref{th: convergence} for those nodes with relatively weak signals.
This property is a distinct advantage of our dynamic regularized method over the existing joint estimation methods \citep{Ma16joint, Cai2016joint}.
Furthermore, define $\hat{S}_{j+}^{(k)} := \hat{S}_{j}^{(k)} \cup \{j\} = \text{supp}\left(\hat{\Theta}_{\cdot,j}^{(k)}\right)$. The following theorem establishes the asymptotic property of nodes with strong signals.

\begin{theorem}[Asymptotic normality]\label{th: asymp}
Assume that all conditions in Theorem \ref{th: supportrecovery} hold, and that $\sqrt n \succ  \max_{j \in \mathcal A}\left( s_j\log p \right)$.
Then, as $n \to \infty$, for every $(j,k) \in \mathcal A \times [K]$, we have
$$
\max_{i \in \hat S_{j+}^{(k)} } ~\sup_{w\in\mathbb R}
\left| \mathbf P\left( \frac{\sqrt {n_k} \left(\hat \Theta^{(k)}_{i,j} - \Theta^{(k)}_{i,j} \right)}{\sigma_{i, \hat S_j}^{(k)}} \le w\right) -\Phi(w) \right| \to 0,
$$
where $\Phi(\cdot)$ is the cumulative distribution function of standard normal distribution and $\sigma_{i, \hat S_j}^{(k)}$ is defined as 
$\left(\sigma_{i, \hat S_j}^{(k)} \right)^2 := \operatorname{Var}\left( \left(\Sigma^{(k)}_{\hat S_{j+}^{(k)}, \hat S_{j+}^{(k)}}\right)^{-1}_{i,\cdot} 
x_{\hat S_{j+}^{(k)}}^{(k)} \left(x_{\hat S_{j+}^{(k)}}^{(k)}\right)^\top
\left(\Sigma^{(k)}_{\hat S_{j+}^{(k)}, \hat S_{j+}^{(k)}}\right)^{-1}_{\cdot, j}
\Big| \hat S_j^{(k)}
\right).$
\end{theorem}

For the strong-signal nodes in \(\mathcal A\), Theorem \ref{th: asymp} shows that the MIGHT method possesses asymptotic normality, thereby enabling statistical inference without the additional debiasing procedure.
This property further strengthens the signal adaptivity of the MIGHT method.
In application, we estimate $\left( \sigma^{(k)}_{i,\hat S_j} \right)^2$ by using
\begin{equation*}
\frac1{n_k} \sum_{\ell=1}^{n_k} \left\{ \left( \hat \Sigma^{(k)}_{\hat S_{j+}^{(k)}, \hat S_{j+}^{(k)}}\right)^{-1}_{i,\cdot} 
		X_{\ell, \hat S_{j+}^{(k)}}^{(k)} \left(X_{\ell, \hat S_{j+}^{(k)}}^{(k)}\right)^\top
		\left(\hat \Sigma^{(k)}_{\hat S_{j+}^{(k)}, \hat S_{j+}^{(k)}}\right)^{-1}_{\cdot, j} \right\}^2
		 - \left( \hat \Theta_{i,j}^{(k)}\right)^2,
\end{equation*}
where we denote by $X_{\ell, \hat S_{j+}^{(k)}}^{(k)} \in \mathbb R^{|\hat S_{j+}^{(k)}|}$ the $\ell$-th observation of the covariates indexed by $\hat S_{j+}^{(k)}$ in the $k$-th dataset.

	\section{Numerical experiments}\label{sec: simu}
	This section evaluates the performance of our proposed method, MIGHT, against five existing methods:
	three joint estimation methods: the joint estimator method by \cite{Guo11joint} (JEM), the group graphical lasso by \cite{Witten13joint} (GGL), and the fast and scalable joint estimator by \cite{wang17joint} (FJEM); and two separate estimation methods: the separate graphical lasso by \cite{FH08} (Sep Glasso) and the separate nodewise $L_0$-regression by \cite{shu2024nodewise} (Sep Node).
	We use the Bayesian Information Criterion (BIC) to select the optimal hyperparameters for these five methods. 
	Specifically, their tuning parameters are chosen to minimize
	$$
	\begin{aligned}
\sum_{k=1}^K &\left\{
-n_k \left( \log \det \left(\hat{\Theta}^{(k)} \right)-\operatorname{tr}\!\left(\hat{\Sigma}^{(k)}\hat{\Theta}^{(k)}\right)\right) \right.\\
&\left. \quad + \|\hat{\Theta}^{(k)}\|_0\left(\log n_k+2\gamma\log p\right)
\right\},
\end{aligned}
$$
where \(\|\hat{\Theta}^{(k)}\|_0\) denotes the number of nonzero entries in the estimated precision matrix. We set \(\gamma=1\), following \cite{barber2015BIC}.
Each tuning parameter is searched over a geometric grid ranging from \(10^{-3}\) to \(1\). 
We upload all code for this section to \url{https://github.com/ShixiangLIUstat/MIGHT-Algorithm}.

We set $p = 100 , K = 10 $ and $n_1 = \cdots = n_K = 100$. 
Data generation begins with an Erd\H{o}s--R\'{e}nyi base graph $G$, where each pair of nodes connects independently with a probability of $0.1$.
We generate $K$ graphs $G_1, \cdots, G_K$ by independently pruning a fraction $\rho $ of edges from the base graph $G$, with $\rho \in (0,1)$ controlling neighbor-similarity and edge-sparsity.
The value of each edge in $ G_k $ is uniformly sampled from $[-1, -0.5] \cup [0.5, 1]$, forming the adjacency matrix $\Omega^{(k)}$. Then, the positive-definite precision matrix $\Theta^{(k)}$ is constructed as $\Theta^{(k)} = \Omega^{(k)} + \left(r + |\Lambda_{\min}(\Omega^{(k)})| \right)\cdot \mathbf I_p$, where parameter $ r > 0 $ modulates the signal strength by inflating diagonal entries, thereby reducing the signal-to-noise ratio (see \eqref{eq: relationship}) in the nodewise regression. 
Finally, by applying the corresponding covariance matrices $\left\{ \Sigma^{(k)} \right\}_{k \in [K]}= \left\{ (\Theta^{(k)})^{-1} \right\}_{k \in [K]}$, we get $K$ datasets as $X^{(1)}, \cdots, X^{(K)}$. 

To assess performance, we use the estimation error in Frobenius norm, defined as $ \sum_{k=1}^K \frac{n_k}{N}\left\| \hat{\Theta}^{(k)} - \Theta^{(k)} \right\|_F$, and the column-wise error in the maximum squared $L_2$-norm (Max L2), defined as $\max_{j \in [p]}  \sum_{k \in [K]} \frac{n_k}{N} \left\| \hat{\Theta}^{(k)}_{\cdot j} - \Theta^{(k)}_{\cdot j} \right\|_2^2$. 
Additionally, we use the Matthews Correlation Coefficient (MCC) to evaluate the selection accuracy of edges in $K$ graphs (MCC-Edge), and of the neighbor set $S_j = \bigcup_{k\in [K]} S_j^{(k)}$ for each node $j \in [p]$ (MCC-Ngbr).

\subsection{Simulation 1: similarity and sparsity}\label{sec 4.1}

We fix $r=0.1$ and vary $\rho \in \{0.2, 0.5, 0.8\}$, with 100 simulation replications for each setting.  
Table \ref{tab: simu1} presents the simulation results; all MCC values have been multiplied by 100, and standard errors are shown in parentheses.
The average runtimes are 0.980s (MIGHT), 2.951s (GGL), 11.170s (JEM), 3.545s (FJEM), 0.976s (Sep Glasso), and 0.529s (Sep Node).
These results indicate that MIGHT is computationally efficient and faster than several joint-estimation competitors, although it is not the fastest method overall. 
Our MIGHT method demonstrates competitive performance across all evaluation metrics compared to competing methods. As $\rho$ increases, the recurrence frequency of shared neighbors across the $K$ graphical models decreases, leading to diminished structural similarity and a decline in the MCC-Ngbr metric.

Appendix \ref{app: simu} provides additional performance metrics (e.g., Hamming Loss and F1 score) and evaluates the robustness of MIGHT under heavy-tailed $t$-distributions.
It also presents extensive simulations across various network structures and parameter settings, including larger \(p=200\), with additional designs that incorporate both structural heterogeneity and value similarity across graphs.

\begin{table}[!t]
\centering
\begin{threeparttable} 
	\caption{The estimation performance across varying $\rho$.}
	\label{tab: simu1}
	\footnotesize
	\setlength{\tabcolsep}{1.2pt}  
	\renewcommand{\arraystretch}{1.2}
	\begin{tabular}{@{}lcccc@{}}  
		\toprule
		& \multicolumn{2}{c}{Estimation Accuracy} & \multicolumn{2}{c}{Structure Recognition} \\
		\cmidrule(lr){2-3} \cmidrule(lr){4-5}
		Methods & Frobenius & Max L2 &
		MCC-Edge & MCC-Ngbr \\
		\midrule
		\multicolumn{2}{c}{ $\rho = 0.2$} & & & \\
		MIGHT      & {\bf 18.979} (0.113) & {\bf 6.257} (0.088)       & {\bf63.662} (0.305) & {\bf 89.673} (0.267) \\
		GGL        & 23.428 (0.111) & 8.772 (0.094)       & 45.276 (0.313) & 39.520 (0.665) \\
		JEM        & 25.272 (0.145) & 9.854 (0.110)       & 54.262 (0.599) & 67.863 (0.531) \\
		FJEM       &49.306 (0.223) & 28.246 (0.155) &
		2.572 (0.333) & 5.310 (0.685) \\
		Sep Glasso & 33.667 (0.151) & 16.061 (0.127)      & 29.715 (0.303) & 42.277 (0.279) \\
		Sep Node   & 24.106 (0.104) & 10.877 (0.135)      & 22.251 (0.198) & 56.403 (0.459) \\ \hline
		\multicolumn{2}{c}{ $\rho = 0.5$} & & & \\
		MIGHT      & {\bf 13.510} (0.127) & {\bf 3.230} (0.075)       & {\bf 71.874} (0.488) & {\bf 87.528} (0.432) \\
		GGL        & 17.704 (0.183) & 5.079 (0.110)       & 56.445 (0.760) & 53.118 (1.872) \\
		JEM        & 18.609 (0.154) & 5.487 (0.099)       & 61.490 (0.767) & 72.527 (0.757) \\
		FJEM       & 39.714 (0.297) & 18.499 (0.368) & 3.140 (0.434) & 5.501 (0.790)\\
		Sep Glasso & 25.434 (0.144) & 9.410 (0.123)       & 40.385 (0.362) & 47.693 (0.366) \\
		Sep Node   & 18.046 (0.108) & 6.092 (0.126)       & 31.874 (0.341) & 62.222 (0.607) \\ \hline
		\multicolumn{2}{c}{ $\rho = 0.8$} & & & \\
		MIGHT      & {\bf 7.290} (0.062)  & {\bf 1.056} (0.027)  & {\bf 81.258} (0.310) & {\bf 83.614} (0.297) \\
		GGL        & 9.610 (0.072)  & 1.577 (0.028)       & 55.902 (0.209) & 58.346 (0.237) \\
		JEM        & 10.210 (0.080) & 1.811 (0.038)       & 69.879 (0.504) & 74.619 (0.562) \\
		FJEM       & 32.331 (2.605) & 652.6 (360.1) 
		& 2.658 (0.293) & 2.617 (0.391)\\
		Sep Glasso & 14.234 (0.076) & 3.157 (0.045)       & 55.920 (0.269) & 51.550 (0.202) \\
		Sep Node   & 9.824 (0.066)  & 1.801 (0.036)       & 54.403 (0.381) & 68.881 (0.430) \\ 
		\bottomrule
	\end{tabular}
\end{threeparttable} 
\end{table}

\subsection{Simulation 2: signal strength}
\begin{figure}
\centering
\includegraphics[width=0.9 \linewidth]{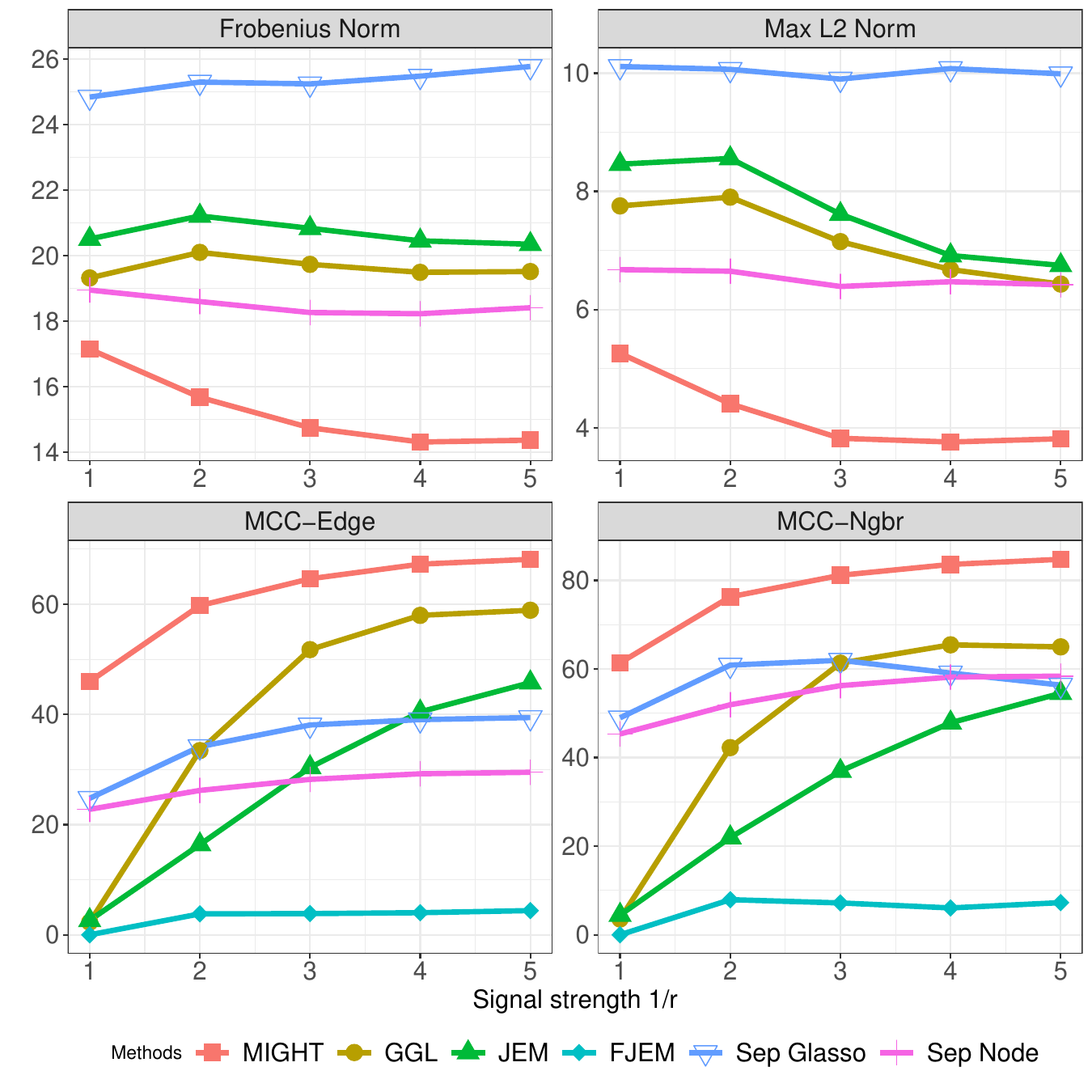}
\caption{Performance metrics with increasing signal strength ($1/r$), with 50 replications for each point. 
	All MCC values have been multiplied by 100.
	The extremely large estimation errors of FJEM are not displayed due to the plotting scale.  
}\label{fig:fig2}
\end{figure}

We fix $\rho=0.5$ and vary $r \in \{ 0.2, 0.25, 0.33, 0.5, 1 \}$, with $1/r \in \{1, 2, 3, 4, 5 \}$ measuring the signal strength. 
The results are presented in Figure \ref{fig:fig2}.
As the signal strength increases, the MIGHT method demonstrates lower estimation error and higher selection accuracy, confirming that enhanced signal strength contributes to better performance, consistent with our theoretical results.

\subsection{Simulation 3: asymptotic distribution}
\begin{figure}
\centering
\includegraphics[width=1\linewidth]{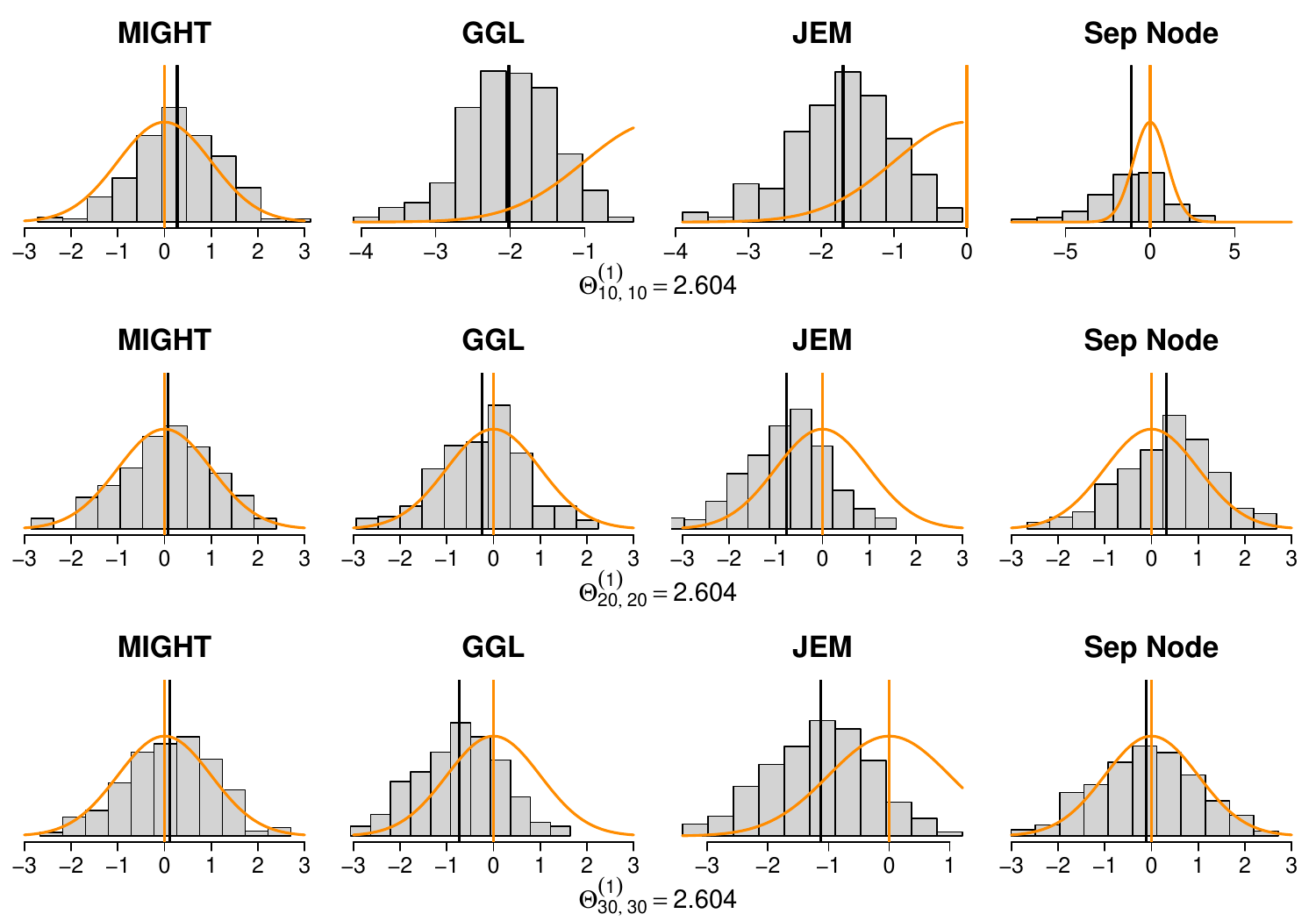}
\caption{Histogram of (diagonal) $\sqrt n \left( \hat \Theta_{ii}^{(k)} -\Theta_{ii}^{(k)} \right) / \hat \sigma_{ii}^{(k)}$ among different methods based on 300 replications. 
	The orange curve represents the density of the standard normal distribution, while the black and orange vertical lines indicate the empirical mean and the population mean (0), respectively.}\label{fig: asy1}
	\end{figure}

\begin{figure}
\centering
\includegraphics[width=1\linewidth]{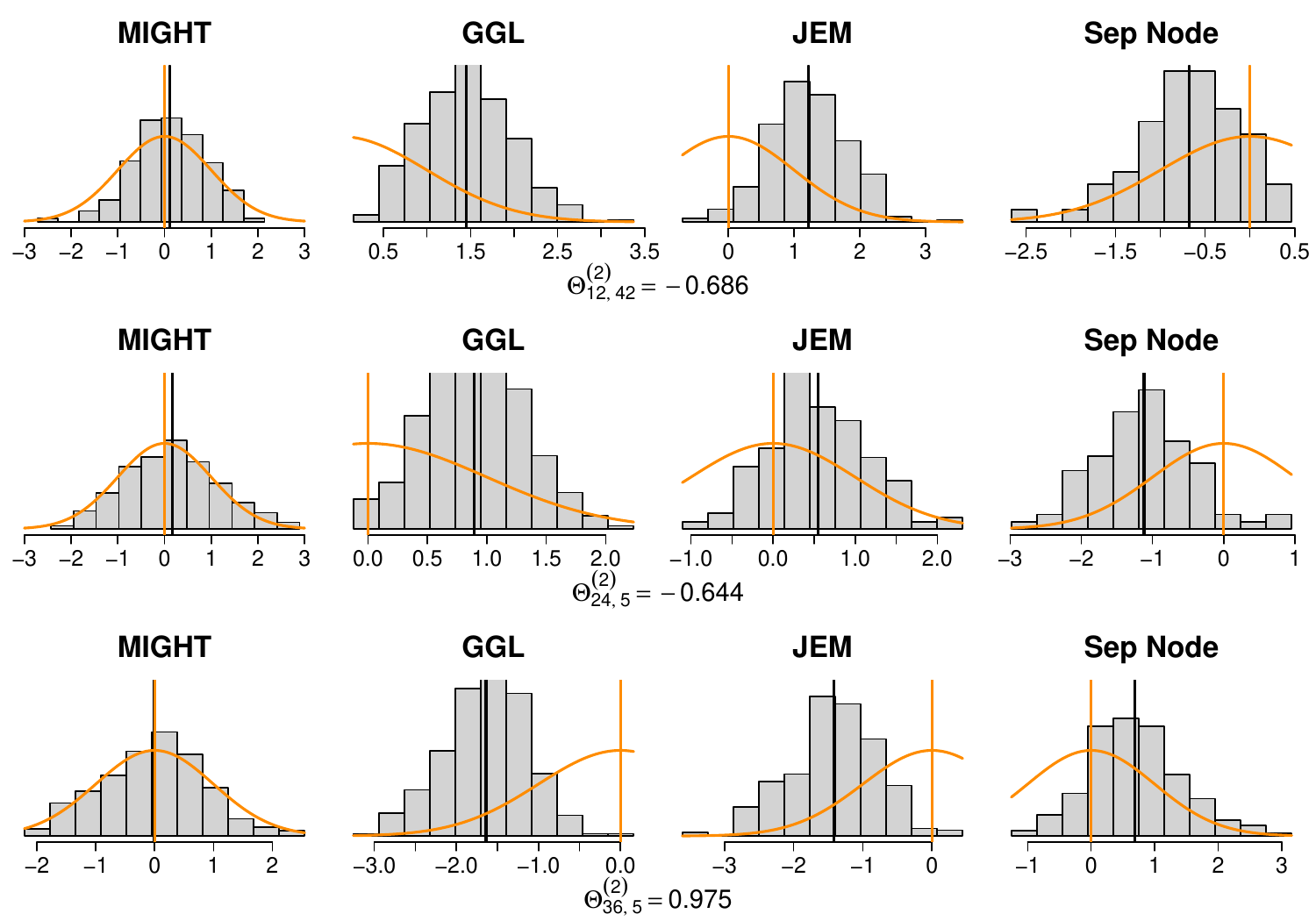}
\caption{Histogram of (off-diagonal) $\sqrt n \left( \hat \Theta_{ij}^{(k)} -\Theta_{ij}^{(k)} \right) / \hat \sigma_{ij}^{(k)}$ among different methods based on 300 replications.}\label{fig: asy2}
\end{figure}

We fix $p=50$, $\rho=0.5$, $r=0.1$, and compare the asymptotic performance of MIGHT, GGL, JEM, and Sep Node methods across six specific matrix entries: three diagonal elements (Figure \ref{fig: asy1}) and three off-diagonal elements (Figure \ref{fig: asy2}). 
The results demonstrate that the MIGHT method achieves better asymptotic normality compared to the separate $\ell_0$ estimator (Sep Node), while GGL and JEM, two joint estimation methods based on M-estimator, fail to exhibit this property. 
	
\section{Real data analysis}\label{sec: real}
In this section, we apply our MIGHT method to a gene-expression cancer RNA-seq dataset, which is accessible via the UCI Machine Learning Repository \citep{gene_expression_cancer_rna-seq_401}.
This dataset encompasses 801 patients and comprises 20,531 genes. 
The patients are categorized into five distinct types based on various tumor types, namely BReast CAncer (BRCA), KIdney Renal cell Carcinoma (KIRC), COlon ADenocarcinoma (COAD), LUng ADenocarcinoma (LUAD), and PRostate ADenocarcinoma (PRAD). The numbers of patients in these subtypes are 300, 78, 146, 141, and 136, respectively. Although these subtypes exhibit biological differences, they all involve tumor tissue, which necessitates joint graphical model learning. 

\subsection{The joint graphical model estimation}
Our primary objective is to generate joint graphical representations of conditional dependence among genes within each type by estimating their precision matrices. Out of the 20,531 genes, we perform a Kruskal-Wallis test on every gene and select the top 500 genes with the lowest $p$-values as a preliminary screening step.
This step is used primarily to reduce the dimension of the original dataset to a computationally manageable set of candidate genes and is separate from the subsequent joint modeling procedure.

\begin{figure}
\centering
\includegraphics[width=1\linewidth]{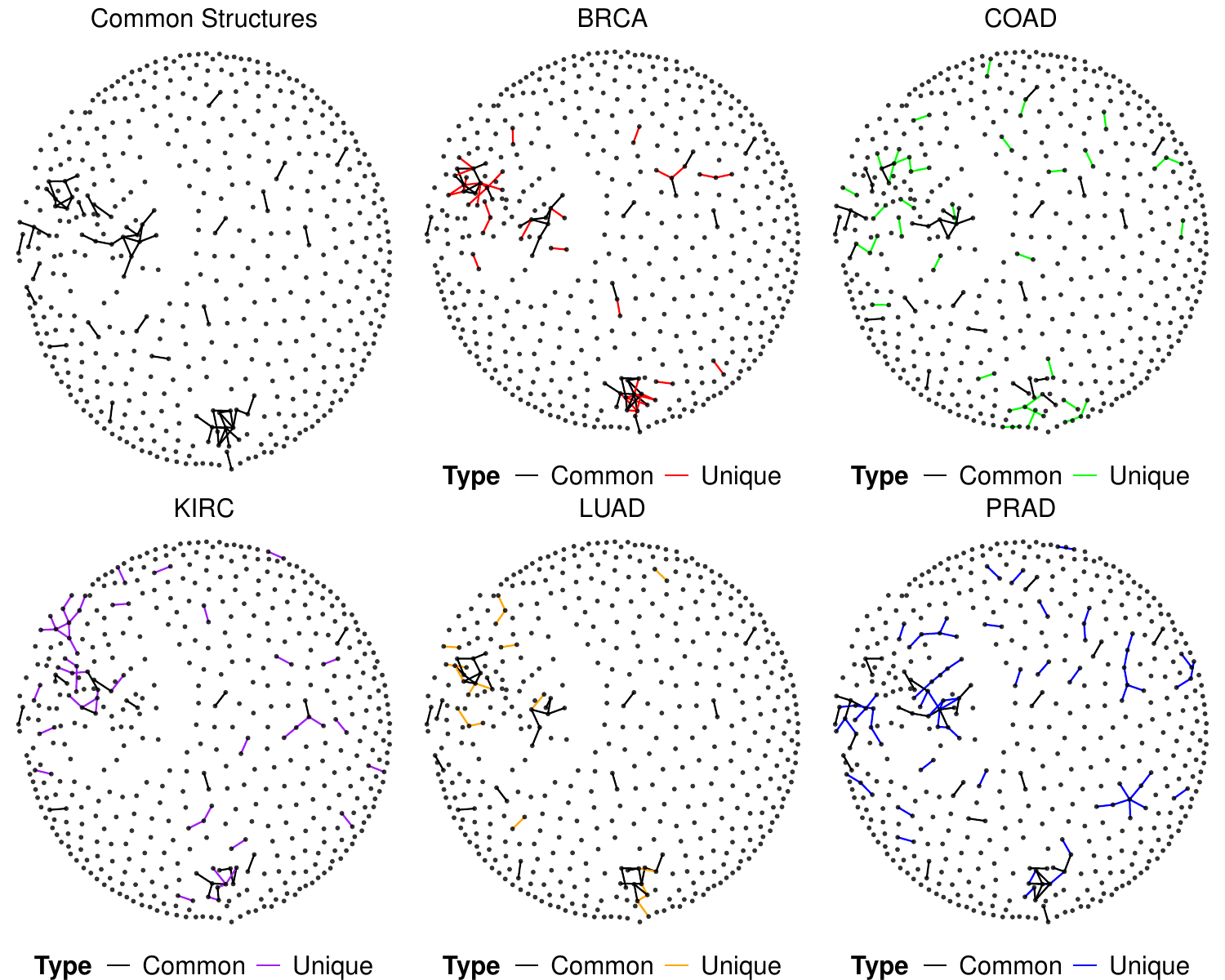}
\caption{Estimated networks for five types of tumors.
	These selected edges characterize conditional dependence on the transformed variables.
}\label{fig:fig_real} 
\end{figure}

Then, to mitigate the non-Gaussian marginal behavior of RNA-seq measurements, we apply a nonparanormal transformation \citep{JMLRliu09} to the selected gene-expression data within each tumor subtype, under a Gaussian-copula interpretation. 
Specifically, this transformation maps the marginal distribution of each gene to an approximately Gaussian scale while preserving the rank information. We implement this procedure using the \texttt{huge.npn} function in the R package \texttt{huge} \citep{zhao2012huge}. 
Such a transformation achieves approximate normality while preserving as much of the raw information as possible, and has therefore been widely adopted in RNA-seq data research \citep{Liang2016bayesian, zhang2021npn, Dong2023JGNsc}.
After this step, the MIGHT method is used to jointly estimate the five subtype-specific precision matrices, which characterize conditional dependence on the transformed (latent Gaussian) scale, with results shown in Figure~\ref{fig:fig_real}.
The black lines are the edges shared by some of the types, and the colored lines are the edges shared by only one specific type.
Notably, most edges appear as black lines, indicating the common structures shared across multiple types of tumors on the transformed scale.

\subsection{The classification task based on QDA}

To further assess the practical utility of the estimated graphical models, we consider a downstream classification task based on quadratic discriminant analysis (QDA), following \cite{Price15JCGS}. 
This task is intended to evaluate the estimated precision matrices as whole objects, rather than to validate support recovery of the joint patterns.
The QDA classifier is constructed using the transformed gene-expression variables, so that the estimated precision matrices and the classification rule are defined on the same scale.
We employ stratified sampling: each time, 80\% of the subjects from each class are randomly selected to form the training set (640 samples), while the remaining samples constitute the testing set (161 samples).
The quadratic discriminant scores are calculated as follows:
\begin{equation}\label{eq: qda}
\delta_k(\mathbf{x})=
\log \hat{\pi}_k + \frac{1}{2}\log \left| \hat \Theta^{(k)} \right|
- \frac12 (\mathbf{x}-\hat{ {\mu}}^{(k)})^\top \hat \Theta^{(k)} (\mathbf{x}-\hat{ {\mu}}^{(k)} ),
\end{equation}
where $k = 1, \cdots, 5$, $\hat{\pi}_k=n_k / N$ is the proportion of observations belonging to the class $k$, and ${\hat\mu}_k = \frac{1}{n_k}\sum_{i=1}^{n_k} \boldsymbol{x}_i^{(k)}$ is the empirical mean for the training set of each class $k$.
The classification rule is then given by $\hat k(\mathbf x) = \arg\max_{k =1, \cdots,5} \delta_k(\mathbf x)$.

The six estimation methods in Section \ref{sec: simu} are used to compute quadratic discriminant scores \eqref{eq: qda} and perform classification predictions, respectively. 
Prediction performance is evaluated through the average TPR, FPR, Accuracy, and MCC. 
Table \ref{tab:real} summarizes the classification results.
The classification based on the MIGHT method yields the best performance in this analysis, illustrating the downstream utility of our method.

\begin{table}[htbp] 
\centering
\begin{threeparttable} 
	\caption{The classification performance for different methods with 50 repetitions.}\label{tab:real}
	\setlength{\tabcolsep}{3pt} 
	\renewcommand{\arraystretch}{1.2}
	\begin{tabular}{@{}lcccc@{}}
		\toprule
		Method & TPR & FPR & Accuracy & MCC \\
		\midrule
		MIGHT      & {\bf0.638} (0.005) & {\bf0.070} (0.001) & {\bf0.739} (0.004) & {\bf0.626} (0.006)  \\
		GGL        & 0.520 (0.004) & 0.110 (0.001) & 0.652 (0.004) & 0.516 (0.005)  \\
		JEM        & 0.620 (0.005) & 0.095 (0.001) & 0.647 (0.005) & 0.531 (0.006)  \\
		FJEM       & 0.184 (0.002) & 0.203 (0.000) & 0.183 (0.008) & -0.007 (0.004) \\
		Sep Glasso & 0.336 (0.003) & 0.162 (0.001) & 0.487 (0.003) & 0.314 (0.004)  \\
		Sep Node   & 0.552 (0.006) & 0.100 (0.002) & 0.596 (0.006) & 0.508 (0.007)  \\
		\bottomrule
	\end{tabular}
\end{threeparttable}  
\end{table}

\section{Conclusion and discussion}\label{sec: conclusion}
This paper proposes an adaptive and computationally efficient method for the joint estimation of multiple graphical models.
Theoretical analysis demonstrates that the proposed method achieves superior estimation accuracy and establishes selection consistency and asymptotic properties.
Both simulation studies and real data analyses demonstrate favorable empirical performance of our method.

The proposed method may have potential applicability to other multi-source dependence-learning problems, such as cross-domain recommendation and multimodal data analysis.
We leave the systematic study of these extensions for future exploration.

\bibliographystyle{unsrtnat} 
\bibliography{bib.bib}       

\clearpage

\appendix
\section{Proof of Proposition \ref{prop: multi-task}}
We acknowledge that Proposition \ref{prop: multi-task} is not a novel result and has been discussed in \cite{Ma16joint, angelini2021jewel}, among many others. However, we briefly outline its derivation here for completeness.
According to the properties of multivariate normality, we have 
\begin{equation}\label{eq: node in app} 
	X_j^{(k)} = X_{\sj}^{(k)} \boldsymbol{\alpha}_{\sj}^{(k)} + \epsilon_j^{(k)} \in \mathbb R^{n_k \times 1},
\end{equation}
where $\epsilon_j^{(k)} \sim N \left( \mathbf 0_{n_k}, \left(\Theta_{j, j}^{(k)}\right)^{-1} \cdot \mathbf I_{n_k} \right)$ is independent of $X_{\sj}^{(k)}$, and 
\begin{equation}\label{eq: alpha in app} 
	\boldsymbol{\alpha}_{\sj}^{(k)} = -\left(\Theta_{j, j}^{(k)}\right)^{-1} \Theta_{\setminus j, j}^{(k)} \in \mathbb R^{p-1}.
\end{equation} 
For each $j \in [p]$, we stack the responses $\left\{ X_j^{(k)} \right\}_{k \in [K]}$, the coefficients $\left\{\boldsymbol{\alpha}_{\sj}^{(k)}\right\}_{k \in [K]}$, and the noise $\left\{ \epsilon_j^{(k)} \right\}_{k \in [K]}$ respectively, and therefore complete the proof of Proposition \ref{prop: multi-task}. 

\section{Proof of Theorem \ref{th: convergence}}
To enhance clarity and conciseness, in the subsequent derivations we set $n_1 = \cdots = n_K =n$, leading $N = nK$. 
For the general case where each observation size $n_k$ varies, the proof remains similar, requiring only the substitution of $1/K$ with the weight $n_k/N$.
We divide the proof into three parts. Appendix \ref{Th1.1} demonstrates that the joint estimation of a fixed column of these precision matrices can be realized as a multi-task learning problem. We also introduce a useful matrix condition (see Definition \ref{EJRIP}) and check that it holds in our model.
Appendix \ref{Th1.2} demonstrates the convergence rate and joint sparse structure in each multi-task learning estimator.
Appendix \ref{Th1.3} integrates the convergence rates and demonstrates the column-wise joint estimation rate over the $K$ graphical models.

For ease of display, define $\Delta_j := \log(s_j K) + (1/s_{0,j}) \log p$.
Denote by $R^{(k)} \in \mathbb R^{p \times p}$ the correlation matrix corresponding to $\Sigma^{(k)}$, and denote by $\Gamma^{(k)}$ the diagonal matrix of $\Sigma^{(k)}$, i.e., $\Gamma^{(k)} = \text{diag}(\Sigma^{(k)}) \in \mathbb R^{p \times p} $.
Define $R :=  \text{diag}(R^{(1)}, \cdots,  R^{(K)})\in \mathbb R^{pK \times pK}$, which is a block-diagonal matrix with each $R^{(k)}$ being one of its diagonal blocks.
The main proof is based on a \textbf{fixed} node $j$, and one can apply a probability union bound to make our results hold for every $j \in [p]$ simultaneously.

\subsection{Preliminaries}\label{Th1.1}

\subsubsection{A scaled regression form}\label{Th1.1.1}

First, we rescale the nodewise regression \eqref{eq: multi-task} so that each column of the covariate has an $\ell_2$ norm of $\sqrt{C_0N}$, where $C_0$ is a positive constant to be determined later. 
Define $Z^{(k)}:= \sqrt{C_0 K}X^{(k)} \left(\hat\Gamma^{(k)}\right)^{-1/2} \in \mathbb R^{n\times p}$, where $\hat \Gamma^{(k)} = \frac1n \text{diag}\left( \left(X^{(k)}\right)^\top X^{(k)} \right)\in \mathbb R^{p\times p}$ is the empirical (diagonal) variance matrix of the $k$-th matrix.
We denote by $ Z_{\setminus j}^{(k)} \in \mathbb R^{n \times (p-1)
}$ the submatrix of $Z^{(k)}$ without its $j$-th column, and define $\bb_{\sj}^{(k)} := \frac1{\sqrt{C_0K}} \left(\hat\Gamma^{(k)}_{\setminus j, \setminus j }\right)^{1/2} \boldsymbol{\alpha}_{\setminus j}^{(k)} \in \mathbb R^{p-1}$, 
{\small
	$$
	Z_{\setminus j} :=  \begin{pmatrix}
		Z_{\setminus j}^{(1)}& \cdots  &\mathbf O_{n\times (p-1)}\\
		\vdots& \ddots& \vdots\\
		\mathbf O_{n\times (p-1)} &\cdots&  Z_{\setminus j}^{(K)}
	\end{pmatrix}, 
	\quad 
	\boldsymbol{\beta}_{\setminus j}:=\begin{pmatrix}
		\bb_{\sj}^{(1)}\\ \vdots \\
		\bb_{\sj}^{(K)}
	\end{pmatrix}.  
	$$
}

\noindent Then the nodewise regression \eqref{eq: multi-task} can be abbreviated as
\begin{equation}\label{eq: rescale}
	X_j = Z_{\setminus j} \boldsymbol{\beta}_{\setminus j} + \epsilon_j,
\end{equation}
and we ensure that each column of $Z_{\setminus j} = \text{diag}\left(Z_{\sj}^{(1)}, \cdots,Z_{\sj}^{(K)}  \right)$ has an $\ell_2$ norm of $\sqrt{C_0 N}$.

\subsubsection{Empirical Joint Restricted Isometry Property condition}\label{Th1.1.3}
For arbitrary positive constants $C_1, C_2$, define the set consisting of all $(C_1 s, C_2 s_0)$-joint sparse support sets as 
{
	\begin{equation*}
		\begin{aligned}
			\mathcal S(C_1s, C_2s_0)
			:=\left\{ S=\left(S^{(1)}, \cdots, S^{(K)} \right) 
			~\left|~
			\begin{aligned}
				& \left| \bigcup_{k \in [K]} S^{(k)} \right|\le C_1 s, \\
				& \sum_{k\in[K]}  | S^{(k)}  |\le C_1 C_2 ss_0
			\end{aligned} \right. \right\}.
		\end{aligned}
	\end{equation*}
}

\noindent For the multi-task learning corresponding to the $j$-th node, define 
\begin{equation}\label{eq: supp j}
	S_j := \left( \text{supp}(\Theta_{\sj,j}^{(1)}), \cdots, \text{supp}(\Theta_{\sj,j}^{(K)})\right).
\end{equation}
By Assumption \ref{assume: sparse}, we learn that $S_j \in \mathcal{S}(s, s_0)$.
Now we introduce a useful matrix condition as \textit{Empirical Joint Restricted Isometry Property (EJRIP)} condition. 
\begin{definition}[Empirical Joint Restricted Isometry Property]\label{EJRIP}
	We say $Z_{\setminus j} \in \mathbb R^{N \times (p-1)K
	}$ satisfies the EJRIP condition $EJRIP(C_1s_j, C_2s_{0,j},\delta)$ if and only if there exists a constant $\delta \in (0,1)$ and two positive constants $C_1, C_2$ (solely determined by $\delta$) such that
	{\small
		\begin{equation}\label{eq: JRIP}
			\begin{aligned}   
				1-\delta \le &\min_{ S\in \mathcal S( C_1 s_j, C_2 s_{0,j} )} \Lambda_{\min} \left(\frac1N \left(Z_{\setminus j}^\top Z_{\setminus j}\right)_{SS}\right)\\
				\le& \max_{ S\in \mathcal S( C_1 s_j, C_2 s_{0,j} )} \Lambda_{\max} \left(\frac1N \left(Z_{\setminus j}^\top Z_{\setminus j}\right)_{SS}\right) \le 1+\delta,
			\end{aligned}
		\end{equation}
	}
	
	\noindent where $\left(Z_{\setminus j}^\top Z_{\setminus j}\right)_{SS}$ is the submatrix of $ Z_{\setminus j}^\top Z_{\setminus j} \in \mathbb R^{(p-1)K \times (p-1)K}$ whose rows and columns are both listed in the joint sparse set $S\in \mathcal S( C_1 s_j, C_2 s_{0,j} )$.
\end{definition}

We next prove that $Z_{\setminus j}$ indeed satisfies the EJRIP$(C_1s_j , C_2s_{0,j},\delta)$ condition \eqref{eq: JRIP} with probability greater than $1- C_3 e^{-C_4(\log K + \log p )}$.
First, given a fixed $j \in [p]$, we check that the restricted eigenvalues of $\frac1N Z_{\setminus j}^\top Z_{\setminus j}$ are bounded.
For ease of display, define abbreviations
\begin{equation*}
	\begin{aligned}
		D_\Sigma^{(k)}  = \hat \Sigma^{(k)} - \Sigma^{(k)},
		\quad  D_\Gamma^{(k)} = ( \hat \Gamma^{(k)} )^{-1/2} - ( \Gamma^{(k)} )^{-1/2}.
	\end{aligned}
\end{equation*}
Then, with probability greater than $1- C_3 e^{-C_4(\log K + \log p )}$, we have 
{\small
	\begin{equation}\label{eq: empirical}
		\begin{aligned}
			&\max_{S \in  \mathcal S( C_1 s_j, C_2 s_{0,j} )}  \max_{\ell \in [|S|]} \left|\Lambda_\ell\left(\frac1N \left(Z_{\setminus j}^\top Z_{\setminus j}\right)_{SS}\right) - \Lambda_\ell\left(C_0 R_{SS} \right) \right|\\
			\le& \max_{S \in  \mathcal S( C_1 s_j, C_2 s_{0,j} )} \left\| \frac1N \left(Z_{\setminus j}^\top Z_{\setminus j}\right)_{SS} - C_0 R_{SS} \right\|_2\\
			\le& C_0 \max_{k,S^{(k)} } \Big\| \Big(  (\hat \Gamma^{(k)} )^{-1/2} \hat \Sigma^{(k)}  (\hat \Gamma^{(k)}  )^{-1/2} \\
			&\qquad \qquad \qquad  - ( \Gamma^{(k)} )^{-1/2} \Sigma^{(k)} ( \Gamma^{(k)} )^{-1/2} \Big)_{S^{(k)} S^{(k)}} \Big\|_2\\
			\le& C_0 \max_{k,S^{(k)} } \left\|  (\hat \Gamma^{(k)}_{S^{(k)} S^{(k)}} )^{-1/2}\right\|_2^2  \left\| \left( D_\Sigma^{(k)} \right)_{S^{(k)} S^{(k)}}  \right\|_2 \\
			&+ C_0 \max_{k,S^{(k)} } \left\|  ( \hat \Gamma^{(k)}_{S^{(k)} S^{(k)}}  )^{-1/2} \right\|_2 
			\left\|\Sigma^{(k)}_{S^{(k)} S^{(k)}} \right\|_2 
			\left\|  \left(D_\Gamma^{(k)}\right)_{S^{(k)} S^{(k)}} \right\|_2\\
			&+C_0 \max_{k,S^{(k)} } \left\|  \left(D_\Gamma^{(k)}\right)_{S^{(k)} S^{(k)}} \right\|_2  \left\|\Sigma^{(k)}_{S^{(k)} S^{(k)}} \right\|_2  \left\| ( \Gamma^{(k)}_{S^{(k)} S^{(k)}} )^{-1/2} \right\|_2  \\
			\le& \frac{C_0}{\sqrt n}\left\{ \frac{2C_5}{\underline C} \sqrt{\log K + s_j \log p} + \frac{(1+\sqrt2) \overline C C_7}{\sqrt{\underline C}} \sqrt{\log(pK)} \right\},
		\end{aligned}
	\end{equation}
}

\noindent where the first inequality follows from Weyl's inequality, and 
$\max_{k,S^{(k)} }$ is the abbreviation of $\max_{k \in [K]}~ \max_{|S^{(k)}|\le C_1 s_j} $.
The last inequality in \eqref{eq: empirical} follows from Lemma \ref{le: eigen} with assuming $n \ge \frac{\overline C C_7^2 \log(pK)}{(\sqrt2 -1)^2}$.
The constants $C_5, C_6, C_7$ come from Lemma \ref{le: eigen} and the constants $\overline C, \underline C$ come from Assumption \ref{assume: eigen}.

Second, we give the upper and lower bounds for the restricted eigenvalues of $R$. By the definition of $R$ and Assumption \ref{assume: eigen}, all eigenvalues of $R$ are well bounded. Hence there exist two constants $\overline C_S = \overline C/ \underline C , \underline C_S = \underline C/\overline C$ leads that
\begin{equation*}
	\begin{aligned}
		C_0 \underline C_S \le& \min_{ S \in \mathcal S( C_1 s_j, C_2 s_{0,j} )} \Lambda_{\min} \left( C_0 R_{SS} \right)\\
		\le& \max_{ S \in \mathcal S( C_1 s_j, C_2 s_{0,j} )} \Lambda_{\max} \left( C_0 R_{SS}\right) \le C_0 \overline C_S.
	\end{aligned}
\end{equation*}
Then, with sufficiently large sample size $n$, from \eqref{eq: empirical} we have 
\begin{equation*}
	\begin{aligned}
		&\max_{S \in  \mathcal S( C_1 s_j, C_2 s_{0,j} )} \max_{\ell \in [|S|]} \left|\Lambda_\ell\left(\frac1N \left(Z_{\setminus j}^\top Z_{\setminus j}\right)_{SS}\right) - \Lambda_\ell\left(C_0 R_{SS} \right) \right|\\
		&\le \frac{ \underline C_S }{\overline C_S + \underline C_S} .
	\end{aligned}
\end{equation*}
Therefore, by taking $\delta \in \left(\frac{\overline C_S}{\overline C_S + \underline C_S} ,1 \right)$ and $C_0 \in \left( \frac{1-\delta}{\underline C_S} + \frac{1}{\overline C_S + \underline C_S}, \frac{\delta}{\overline C_S} + \frac{1}{\overline C_S + \underline C_S} \right)$, we have
\begin{equation}\label{eq: ejrip zj}
	\begin{aligned}
		1 - \delta \le &\min_{ S \in \mathcal S( C_1 s_j, C_2 s_{0,j} )} \Lambda_{\min} \left(\frac1N \left(Z_{\setminus j}^\top Z_{\setminus j}\right)_{SS}\right)\\
		\le & \max_{ S \in \mathcal S( C_1 s_j, C_2 s_{0,j} )} \Lambda_{\max} \left(\frac1N \left(Z_{\setminus j}^\top Z_{\setminus j}\right)_{SS}\right) \le 1+\delta .
	\end{aligned}
\end{equation}
Finally, by using a probability union bound over all $j \in [p]$, we conclude that, with probability greater than $1- C_3 e^{-C_4(\log K + \log p )}$, the corresponding $Z_{\setminus j}$ satisfies the EJRIP condition \eqref{eq: JRIP} for every $j \in [p]$.

\subsection{Dynamic-regularized regression}\label{Th1.2}
For a fixed $j\in[p]$, Appendix \ref{Th1.1} shows that the rescaled design matrix $Z_{\setminus j}$ satisfies the EJRIP$(C_1s_j, C_2s_{0,j},\delta)$ condition with probability greater than $1- C_3 e^{-C_4(\log K + \log p )}$, and the corresponding noise $\epsilon_j \in \mathbb R^N$ is i.i.d. sub-Gaussian, with its sub-Gaussian parameter $\sigma_j^2 := \max _{k \in [K]} \left(\Theta_{jj}^{(k)} \right)^{-1} \le \overline C$.
And we also get an upper bound of $\|  \bb_{\sj}\|_2 $ as 
\begin{equation*}
	\begin{aligned}
		\|\bb_{\sj} \|_2  \le& \sqrt{\sum_{k \in [K]} \frac1{C_0 K} \left\| \hat \Gamma_{\sj,\sj}^{(k)} \right\|_2 \frac{\|\Theta_{\sj,j}^{(k)} \|_2^2 }{\left( \Theta_{j,j}^{(k)} \right)^2 } }\\
		\le& \sqrt{\sum_{k \in [K]} \frac{2 \overline C^3}{C_0 K \underline C^2} }
		~=~ \sqrt{ \frac{\overline C^2 \left(\overline C^2 + \underline C^2 \right) }{\underline C^3} }.
	\end{aligned}
\end{equation*}
Thus, by following Theorem 1 in \cite{liu2025exact} and taking a sufficiently large initial threshold parameter $\lambda_0 = C_2^{\text{alg}} \ge \sqrt{ \frac{\overline C^2 (\overline C^2 + \underline C^2 ) }{\underline C^3} }$, we learn that the following holds for every iteration round $t\ge 0$ in the dynamic stage of Algorithm \ref{IHT}:
\begin{equation}\label{eq: pattern}
	\begin{aligned}
		\|\hat \bb_{\sj}^t - \bb_{\sj} \|_2 \le& \sqrt{A ss_0 } \lambda_{t}~ , \\
		\text{supp}(\hat \bb_{j}^{t}) \in& \left( (1+C_1')s_j, \frac{1+2C_1'}{1+ C_1'}s_{0,j} \right),
	\end{aligned}
\end{equation}
where we take $\lambda_t = (\kappa^t \lambda_0) \vee \lambda_{\infty,j}$, $C_1'=  \frac{8}{\left( \kappa/\delta -1 \right)^2}$, $\lambda_{\infty,j} = \sqrt{ C_\infty \cdot \frac{12 \overline C (\log K + \frac{\log p}{s_{0,j}})}{N}}$, and 
\begin{equation*}
	C_\infty = \frac{4\left( \kappa/\delta+\sqrt3-1 \right)^2}{\left( \kappa/\delta -1 \right)^2},
	\quad A =  C_1' \left( \frac\kappa\delta\right)^2 \left( 1-\sqrt{\frac1{C_\infty}}\right)^2 .
\end{equation*}

Assume that $\lambda_t$ reaches $\lambda_{\infty,j}$ in the $t_0$-th iteration, we then begin the (fixed-stage) iteration with a fixed thresholding parameter $ \lambda_{\text{fix},j} = C_\text{fix}  \sqrt{ \frac{ \log(\hat s_j K)+ (1/s_{0,j}) \log p }{N}}$, where $\hat s_j = |\hat G_j^{t_0}| \vee 1$ and $\hat G_j^{t_0} \subset [p]\setminus\{j\}$ is the index set of the discovered covariates in the $t_0$-th iteration.
Following the similar induction and techniques in Theorem 1 from \cite{liu2025exact}, with a sufficiently large constant $C_\text{fix}$, we still make sure that, with a probability greater than $1- O\left( e^{-C_4 \log (pK) }\right)$, $\hat \bb_{\sj}^t$ is still a $\left( (1+C_1')s_j, \frac{1+2C_1'}{1+ C_1'}s_{0,j} \right)$-sparse estimator and $ \| \hat\bb_{\sj}  -  \bb_{\sj} \|_2 \lesssim \sqrt{s_j s_{0,j}} \lambda_{\text{fix},j} $.
Hence, by using a constant $C_D>0$ (determined by $\delta$ and $\kappa$) and following a probability union bound over all $ j \in [p]$, for every $j \in [p]$, the output of Algorithm \ref{IHT} satisfies
\begin{equation}\label{eq: part2}
	\left\| \hat\bb_{\sj}  -  \bb_{\sj} \right\|_2^2 \le C_{D}\times  \frac{ s_j s_{0,j}\log(s_jK) + s_j \log p }{N},
\end{equation}
with a probability greater than $1- O\left( e^{-C_4 \log (pK) }\right)$. 

\subsection{Integration}\label{Th1.3}

This part focuses on the joint estimation rate. 
We separate the final result into two parts: the estimation error on the diagonal $\sum_{k \in [K]} \left( \hat \Theta_{jj}^{(k)} - \Theta_{jj}^{(k)}\right)^2 $ and the off-diagonal $\sum_{k \in [K]} \left\| \hat \Theta_{\sj,j}^{(k)} - \Theta_{\sj,j}^{(k)}\right\|_2^2 $.

\subsubsection{Preliminaries}\label{Th1.3.1}

We separate the estimator $\hat\bb_{\sj}$ into $( \hat\bb_{\sj}^{(1)}, \cdots , \hat\bb_{\sj}^{(K)} )$. Then define 
{\small
	\begin{align*}
		(\sigma_j^{(k)})^2 &: = \frac1n \mathbf E \left\| X_j^{(k)} - X_{\sj}^{(k)} \boldsymbol\alpha_{\sj}^{(k)} \right\|_2^2 = \left( \Theta_{jj}^{(k)} \right)^{-1} ,\\
		(\hat \sigma_j^{(k)})^2 &: =  \frac1n \left\| X_j^{(k)} - X_{\sj}^{(k)} \hat{\boldsymbol\alpha}_{\sj}^{(k)} \right\|_2^2 
		= \frac1n \left\| X_j^{(k)} - Z_{\sj}^{(k)} \hat \bb_{\sj}^{(k)} \right\|_2^2 ,
	\end{align*}
}

\noindent and take $\hat \Theta_{jj}^{(k)} = \left(\hat \sigma_j^{(k)} \right)^{-2}$.
With a probability greater than $1- O\left( e^{-C_4 \log (pK) }\right)$ it yields that
{ 
	\begin{equation*}
		\begin{aligned}
			&\left| (\sigma_j^{(k)})^2 - (\hat \sigma_j^{(k)})^{2}\right|\\
			\le & \left| \frac1n \| \epsilon_j^{(k)}\|_2^2 - (\sigma_j^{(k)})^2 \right| 
			+ \frac1n \left\| Z_{\sj}^{(k)} \left(\bb_{\sj}^{(k)}- \hat \bb_{\sj}^{(k)} \right) \right\|_2^2\\
			&+ \frac2n \left \langle \left(Z_{\sj}^{(k)}\right)^\top \epsilon_j^{(k)}, ~\bb_{\sj}^{(k)}- \hat \bb_{\sj}^{(k)} \right \rangle \\
			\le & 4(\sigma_j^{(k)})^2 \sqrt{\frac{C_4 \log(pK)}{n}} 
			+\frac{2N}{n}\left\| \bb_{\sj} - \hat \bb_{\sj} \right\|_2^2\\
			&+ \frac{6N}{n} \sqrt{\frac{12\overline C C_1's_j s_{0,j} \Delta(s_{0,j} ) }{ N} } \left\| \bb_{\sj} - \hat \bb_{\sj} \right\|_2 \\
			\lesssim&  \sqrt{\frac{ \log(pK)}{n}} 
			+ \frac{s_j s_{0,j} \log(s_j K) + s_j \log p}{n},
		\end{aligned}
	\end{equation*}
}

\noindent where the second inequality follows from Lemma \ref{le: chi}, Lemma  \ref{le: control} and EJRIP condition, and the final inequality follows from \eqref{eq: part2} and the following: 
{\small
	\begin{equation*}
		\underline C 
		\le \Lambda_{\max}^{-1} \left( \Theta^{(k)} \right)
		\le (\sigma_j^{(k)})^2 = \left( \Theta_{jj}^{(k)} \right)^{-1}
		\le \Lambda_{\min}^{-1} \left( \Theta^{(k)} \right)
		\le \overline C.
	\end{equation*}
}

\noindent Therefore, with sufficiently large sample size $n$, we get $(\hat \sigma_j^{(k)})^{2} \in \left( \underline C/2, \overline C +\underline C/2 \right)$.

\subsubsection{Diagonal}\label{Th1.3.2}
By Appendix \ref{Th1.3.1}, for every $k \in [K]$ we get
{\small
	\begin{equation*}
		\left( \hat \Theta_{jj}^{(k)} - \Theta_{jj}^{(k)}\right)^2
		=  \left(  \frac{(\hat\sigma_j^{(k)})^2-(\sigma_j^{(k)})^2}{(\hat \sigma_j^{(k)} \sigma_j^{(k)})^2} \right)^2
		\asymp \left( (\hat\sigma_j^{(k)})^2-(\sigma_j^{(k)})^2 \right)^2.
	\end{equation*}
}

\noindent Then, with a probability greater than $1- O\left( e^{-C_4 \log (pK) }\right)$, we integrate the diagonal estimation error as: 
{\small
	\begin{equation}\label{eq: diagonal}
		\begin{aligned}
			&\sum_{k \in [K]}\left( \hat \Theta_{jj}^{(k)} - \Theta_{jj}^{(k)}\right)^2
			\asymp \sum_{k \in [K]} \left( (\hat\sigma_j^{(k)})^2-(\sigma_j^{(k)})^2 \right)^2\\
			\lesssim& \sum_{k \in [K]} \left( \frac1n \| \epsilon_j^{(k)}\|_2^2 - (\sigma_j^{(k)})^2 \right)^2 
			+ \frac1{n^2} \sum_{k \in [K]} \left\| Z_{\sj}^{(k)} \left(\bb_{\sj}^{(k)}- \hat \bb_{\sj}^{(k)} \right) \right\|_2^4\\
			&+ \frac{N^2}{n^2}  \sum_{k \in [K]}
			\left\| \frac1N \left( (Z_{\sj}^{(k)})^\top \epsilon_j^{(k)} \right)_{\text{supp}(\bb_{\sj}^{(k)}- \hat \bb_{\sj}^{(k)})} \right\|_2^2 \left\|\bb_{\sj}^{(k)}- \hat \bb_{\sj}^{(k)} \right\|_2^2  \\
			\lesssim & \frac{K + \log(pK)}{n}  + \left(\frac{s_j s_{0,j} \log(s_j K) + s_j \log p}{n} \right)^2,
		\end{aligned}
	\end{equation}
}

\noindent where the last inequality follows from Lemma \ref{le: chi} (with taking $q=K, t=C_4 \log(pK)$), and also follows from the EJRIP condition, \eqref{eq: part2} and Lemma \ref{le: control}.

\subsubsection{Off-diagonal}\label{Th1.3.3}
By the rescaling formulation in Appendix \ref{Th1.1.1}, we get
\begin{align*}
	\Theta_{\sj,j}^{(k)} &= -\Theta_{j,j}^{(k)} \boldsymbol\alpha_{\sj}^{(k)},  
	\quad \boldsymbol{\alpha}_{\setminus j}^{(k)} = \sqrt{C_0K}  \left(\hat\Gamma^{(k)}_{\setminus j, \setminus j }\right)^{-1/2}  \bb_{\sj}^{(k)},\\
	\hat \Theta_{\sj,j}^{(k)} &= - \hat\Theta_{j,j}^{(k)} \hat{\boldsymbol\alpha}_{\sj}^{(k)} , 
	\quad \hat{\boldsymbol{\alpha}}_{\setminus j}^{(k)} = \sqrt{C_0K}  \left(\hat\Gamma^{(k)}_{\setminus j, \setminus j }\right)^{-1/2}  \hat \bb_{\sj}^{(k)}.
\end{align*}
Therefore, with a probability greater than $1- O\left( e^{-C_4 \log (pK) }\right)$, we have
{\small
	\begin{equation}\label{eq: offdiagonal}
		\begin{aligned}
			&\sum_{k \in [K]} \left\| \hat \Theta_{\sj,j}^{(k)} - \Theta_{\sj,j}^{(k)} \right\|_2^2 \\
			=& C_0K \sum_{k \in [K]} \left\| \left(\hat\Gamma^{(k)}_{\setminus j, \setminus j }\right)^{-1/2}  \hat \bb_{\sj}^{(k)} \hat\Theta_{j,j}^{(k)} -  \left(\hat\Gamma^{(k)}_{\setminus j, \setminus j }\right)^{-1/2}  \bb_{\sj}^{(k)}\Theta_{j,j}^{(k)} \right\|_2^2 \\
			\lesssim& K \sum_{k \in [K]} \left\| \left(\hat\Gamma^{(k)}_{\setminus j, \setminus j }\right)^{-1/2}  \left( \hat \bb_{\sj}^{(k)} - \bb_{\sj}^{(k)} \right)\hat\Theta_{j,j}^{(k)}\right\|_2^2\\
			& + K \sum_{k \in [K]}\left\|\left(\hat\Gamma^{(k)}_{\setminus j, \setminus j }\right)^{-1/2}  \bb_{\sj}^{(k)} \left( \hat\Theta_{j,j}^{(k)}-\Theta_{j,j}^{(k)} \right) \right\|_2^2\\
			\overset{(i)}\lesssim& K \left\| \hat \bb_{\sj} - \bb_{\sj} \right\|_2^2
			+ \sum_{k \in [K]}\left\| \Theta_{\sj,j}^{(k)}\right\|_2^2 \left(\Theta_{jj}^{(k)} \right)^{-2} \left( \hat\Theta_{j,j}^{(k)}-\Theta_{j,j}^{(k)} \right)^2 \\
			\lesssim&  \frac{K+  s_j s_{0,j} \log(s_j K) + s_j \log p }{n},
		\end{aligned}
	\end{equation}
}

\noindent where inequality (i) holds because $\left\| \left(\hat\Gamma^{(k)}_{\setminus j, \setminus j }\right)^{-1/2} \right\|_2$ and $\hat\Theta_{jj}^{(k)}$ are both upper bounded by constants with sufficiently large sample size (concluded from Lemma \ref{le: eigen} and Appendix \ref{Th1.3.1} respectively).
The last inequality in \eqref{eq: offdiagonal} follows from that $\left\| \Theta_{\sj,j}^{(k)}\right\|_2 \le \left\| \Theta_{\cdot,j}^{(k)}\right\|_2 \le \Lambda_{\max}(\Theta^{(k)})\le 1/\underline C$, \eqref{eq: part2}, \eqref{eq: diagonal} and $n \gtrsim s_j s_{0,j} \Delta_j $.

\subsubsection{Conclusions}\label{Th1.3.4}
Finally, by combining \eqref{eq: diagonal}, \eqref{eq: offdiagonal}, and the probability union bound, with a probability greater than $1- O\left( e^{-C_4 \log (pK) }\right)$, for every $j \in [p]$ we have 
$$
\sum_{k \in [K]} \left\|\hat \Theta^{(k)}_{\cdot j} - \Theta^{(k)}_{\cdot j} \right\|_2^2 
\lesssim  \frac{ K+  s_j s_{0,j} \log(s_j K) + s_j \log p}{n},
$$
which proves the first result in Theorem \ref{th: convergence}.

Notably, $\left( \hat \Theta^{(1)}_{\sj, j}, \cdots, \hat \Theta^{(K)}_{\sj, j}\right)$ is obtained by rescaling $\hat \bb_{\sj}$ through scalar multiplication, and hence preserving the same sparsity pattern \eqref{eq: pattern}.
Consequently, the joint estimation enjoys a joint sparse structure, which proves the second result in Theorem \ref{th: convergence}.

\section{Proof of Theorem \ref{th: supportrecovery}}
Since $\left( \hat \Theta^{(1)}_{\sj, j}, \cdots, \hat \Theta^{(K)}_{\sj, j}\right)$ preserves the same sparsity pattern of $\hat \bb_{\sj}$, we only need to prove that the multi-task regression (11) achieves exact support recovery when the node $j$ satisfies Assumption \ref{assume: betamin}.
The analysis of the dynamic iteration stage in Algorithm \ref{IHT} is the same as in Appendix \ref{Th1.2}.
In the fixed iteration stage, we set the thresholding $ \lambda_{\text{fix},j}  = C_\text{fix} \sqrt{ \frac{ \log(\hat s_j K) + (1/s_{0,j}) \log p )}{N}}$ (where $\hat s_j = |\hat G_j^{t_0}| \vee 1$).
In the case $\min_{i \in G_j} \sum_{k\in[K]} \left(\bb_{\sj}^{(i,k)}\right)^2 \ge  2C_D \cdot \frac{s_{0,j} \log(s_j K)+ \log p}{N}$, by (19) and a contradiction argument we conclude that $\mathbf P\left\{ s_j/2\le \hat s_j \le (1+C_1')s_j \right\} \ge 1- O\left( e^{-C_4(\log K + \log p )}\right)$.
Consequently, with the same probability, we obtain
$$
\lambda_{\text{fix},j}  \asymp \sqrt{ \frac{ \log( s_j K) + (1/s_{0,j}) \log p )}{N}}.
$$

Following Proposition 2.1 of \cite{liu2025exact}, under signal Assumption 4, the fixed-stage estimator $\hat\bb_{\sj}^{t}$ can be made {\bf arbitrarily} close to the oracle estimator $\tilde \bb_{\sj}$ by taking sufficiently many iterations.
And the oracle estimator $\tilde\bb_{\sj}$ satisfies 
\begin{equation}\label{eq: oraclenodewise}
	\left(\tilde\bb_{\sj}\right)_{S_j} = \left[(Z_{\sj}^\top Z_{\sj})_{S_j,S_j}\right]^{-1}  \left((Z_{\sj})_{S_j}\right)^\top X_j \in \mathbb R^{|S_j|},
\end{equation}
and $\left(\tilde\bb_{\sj}\right)_{S_j^c} =\mathbf 0$. 

Therefore, by setting the final output of Algorithm 1 as $\hat \bb_{\sj} = \hat\bb_{\sj}^{t}$ with arbitrary $ t \gtrsim t_0 +\log N$, we conclude that
\begin{equation}\label{eq: delta}
	\left\| \hat\bb_{\sj}^{t} - \tilde \bb_{\sj} \right\|_2 \le \frac1N.
\end{equation}
Additionally, conditioned on $Z_{\sj}$, $ \left( \tilde \bb_{\sj} - \bb_{\sj} \right)_{S_j} = \left[(Z_{\sj}^\top Z_{\sj})_{S_j,S_j}\right]^{-1}  \left((Z_{\sj})_{S_j}\right)^\top \epsilon_j \in \mathbb R^{|S_j|}$ is entrywise sub-Gaussian with its sub-Gaussian parameter $\overline C \sqrt{ \left\| \left[(Z_{\sj}^\top Z_{\sj})_{S_j,S_j}\right]^{-1} \right\|_2  } $, which leads that
\begin{equation}\label{eq: oracle max}
	\begin{aligned}
		&\mathbf P\left( \| \tilde \bb_{\sj}  - \bb_{\sj} \|_\infty \ge \lambda_{\text{fix},j}/2 \right)\\
		\le & \mathbf{P}\left( \| \tilde \bb_{\sj}  - \bb_{\sj} \|_\infty \ge \lambda_{\text{fix},j}/2 ~\left| ~Z_{\setminus j} \text{ satisfies EJRIP} \right. \right)\\
		& +  \mathbf{P}\left( Z_{\setminus j} \text{ violates EJRIP} \right)\\
		\lesssim& s_j s_{0,j} \cdot\exp \left(- \frac{ \lambda_{\text{fix},j}^2 N (1-\delta)}{8 \sigma^2} \right) + \exp\big(-C_4' \log(pK) \big) \\
		\lesssim& \exp\left\{ -C_4'' \left(\frac{\log p}{s_{0,j}}+ \log(sK) \right) \right\},
	\end{aligned}
\end{equation}
where the last inequality holds by taking the constant $C_\text{fix} \ge 4 \sqrt{\frac{3 \overline C}{1-\delta}}$.

Combining with \eqref{eq: delta} and \eqref{eq: oracle max}, we obtain
$$
\| \hat\bb_{\sj}^{t} - \bb_{\sj} \|_\infty 
\le \| \hat\bb_{\sj}^{t} - \tilde \bb_{\sj} \|_2 + \| \tilde \bb_{\sj}  - \bb_{\sj} \|_\infty
\le  \lambda_{\text{fix},j}.
$$
Therefore, if the signal satisfies $\min_{ k \in [K]}  \min_{i\in S_j^{(k)}} \left|\Theta^{(k)}_{i,j} \right| \ge \lambda_{\text{fix},j}$, our MIGHT estimator achieves the exact support recovery for the fixed $j \in [p]$.
Furthermore, by (17) and a technique similar to Lemma 2, with a probability greater than $1 - C_5 e^{-C_6\log (pK)}$, we have
\begin{equation*}
	\begin{aligned}
		\left\| \tilde \bb_{\sj} - \bb_{\sj} \right\|_2^2 
		\le& \left\| \left(\frac{Z_{\sj}^\top Z_{\sj}}N \right)_{S_j,S_j}^{-1} \right\|_2^2 \cdot \left\| \left( \frac{1}{ N} Z_{\setminus j}^\top \epsilon_j \right)_{S_j} \right\|_2^2 \\
		\lesssim& \frac{ s_j s_{0,j} + \log (pK)}N.
	\end{aligned}
\end{equation*}
Combining this result with \eqref{eq: delta}, with a probability greater than $1 -  O\left( e^{-C_{4} \left(\frac{\log p}{s_{0,j}}+ \log(s_jK) \right) } \right)$ we establish that 
\begin{equation}\label{eq: sharp error}
	\left\| \hat \bb_{\sj} - \bb_{\sj} \right\|_2^2 \lesssim \frac{ s_j s_{0,j} + \log(pK)}N .
\end{equation}
Substituting \eqref{eq: sharp error} into (20) and (21), we derive the tighter error bound:
\begin{equation*}
	\begin{aligned}
		&\mathbf P\left( \sum_{k \in [K]} \left\|\hat \Theta^{(k)}_{\cdot j} - \Theta^{(k)}_{\cdot j} \right\|_2^2 \lesssim \frac{K+s_j s_{0,j} + \log p}{n}  \right) \\
		\ge &1 -  O\left( e^{-C_{4} \left(\frac{\log p}{s_{0,j}}+ \log(s_jK) \right) }  \right).
	\end{aligned}
\end{equation*}

Recall $\mathcal A :=\{j\in [p]: \text{ node j satisfies Assumption 4} \}$, and assume that there exists a constant $\gamma \in (0,1)$ such that $\min_{j \in \mathcal A} \frac{\log p}{s_{0,j}} + \log(s_j K) \ge \gamma \log |\mathcal A|$.
Then, by using the probability union bound, we prove that above results hold for every $j \in \mathcal A$, with a probability at least $1- C_1e^{-C_{2,\gamma} \min_{j \in \mathcal A}\left( \frac{\log p}{s_{0,j}} + \log(s_jK) \right)}$, which completes the proof of Theorem \ref{th: supportrecovery}.

\section{Proof of Theorem \ref{th: asymp}}
The proof of Theorem \ref{th: asymp} is separated into two parts. In Appendix \ref{th3.1}, we analyze the difference between our estimator and the oracle nodewise estimator.
Appendix \ref{th3.2} then demonstrates the asymptotic normality.
Table \ref{tab: asymp} introduces some symbols and their meanings used in the proof of Theorem \ref{th: asymp}.

\begin{table*}[t]
	\caption{Symbols and their meanings.}
	\label{tab: asymp}
	\renewcommand\arraystretch{1.5}
	\resizebox{\linewidth}{!}{ \begin{tabular}{@{}ll@{}}
			\hline
			Symbol   & Meaning \\
			\hline
			$\hat \bb_{\sj} \in \mathbb R^{(p-1)K}$ &The estimator output by Algorithm 1. \\  
			$\hat \bb_{\sj}^{(k)} \in \mathbb R^{p-1}$ &The nodewise estimator for the $k$-th dataset. \\
			$\hat{\boldsymbol{\alpha}}_{\sj}^{(k)} = \sqrt{C_0 K} \left(\hat \Gamma_{\sj,\sj}^{(k)}\right)^{-1/2}\hat \bb_{\sj}^{(k)}$& The nodewise estimator for the $k$-th dataset, after rescaling.\\ 
			$\hat \Theta_{jj}^{(k)} = \left( \frac1n \left\| X_j^{(k)} - X_{\sj}^{(k)}\hat{\boldsymbol{\alpha}}_{\sj}^{(k)} \right\|_2^2 \right)^{-1}$ & The estimation of $\Theta_{jj}^{(k)}$ for the $k$-th dataset.\\ 
			$\hat \Theta_{\sj, j}^{(k)} = -\hat \Theta_{jj}^{(k)} \hat{\boldsymbol{\alpha}}_{\sj}^{(k)} \in \mathbb R^{p-1}$ & The estimation of $\Theta_{\sj j}^{(k)}$ for the $k$-th dataset.\\ 
			\hline
			$ \tilde\bb_{S_j^{(k)}}^{(k)}  = \left[\left(Z_{S_j^{(k)}}^{(k)}\right)^\top Z_{S_j^{(k)}}^{(k)} \right]^{-1} \left(Z_{S_j^{(k)}}^{(k)}\right)^\top X_j^{(k)} $ & The oracle nodewise estimator \eqref{eq: oraclenodewise} for the $k$-th dataset. \\ 
			$\tilde{\boldsymbol{\alpha}}_{S_j^{(k)}}^{(k)} = \left( \hat \Sigma_{S_j^{(k)},S_j^{(k)}}^{(k)}\right)^{-1} \hat \Sigma_{S_j^{(k)},j}^{(k)} \in \mathbb R^{|S_j^{(k)}|}$& The oracle nodewise estimator for the $k$-th dataset, after rescaling.\\
			$\tilde \Theta_{jj}^{(k)} = \left( \frac1n \left\| X_j^{(k)} - X_{S_j^{(k)}}^{(k)} \tilde{\boldsymbol{\alpha}}_{S_j^{(k)}}^{(k)} 
			\right\|_2^2 \right)^{-1}$ & The oracle nodewise estimation of $\Theta_{jj}^{(k)}$ for the $k$-th dataset.\\ 
			$\tilde \Theta_{S_j^{(k)}, j}^{(k)} = -\tilde \Theta_{jj}^{(k)} \tilde{\boldsymbol{\alpha}}_{S_j^{(k)}}^{(k)} \in \mathbb R^{|S_j^{(k)}|}$ & The oracle nodewise estimation of $\Theta_{S_j^{(k)}, j}^{(k)}$ for the $k$-th dataset.\\ 
			\hline
			$\delta_{\sj} = \hat \bb_{\sj} - \tilde \bb_{\sj}~,~~ \delta_{\sj}^{(k)} = \hat \bb_{\sj}^{(k)} - \tilde \bb_{\sj}^{(k)}$ &Difference between our estimator and the oracle nodewise estimator.\\   
			\hline
	\end{tabular} } 
\end{table*}

Recall that
\begin{equation}\label{eq: bjk}
	\begin{aligned}
		\|\bb_{\sj}^{(k)} \|_2 
		\le& \frac{1}{\sqrt{C_0 K}} \left\| \left(\hat \Gamma_{\sj,\sj}^{(k)}\right)^{1/2} \right\|_2 \|\boldsymbol{\alpha}_{\sj}^{(k)} \|_2\\
		\le& \frac{\sqrt{2\overline C } \|\Theta_{\sj,j}^{(k)} \|_2 }{ \Theta_{j,j}^{(k)}  \sqrt{C_0 K}} 
		\le \frac{\sqrt2~ \overline C^{3/2}}{ \underline C \sqrt{C_0 K}}.
	\end{aligned}
\end{equation}
The proof of Theorem \ref{th: asymp} relies on the results of Theorem \ref{th: supportrecovery}, which hold with probability at least $1- C_1e^{-C_{2,\gamma} \min_{j \in \mathcal A}\left( \frac{\log p}{s_{0,j}} + \log(s_jK) \right)}$. Therefore, all subsequent derivations are conditioned on this event. For the sake of brevity, we omit explicit reference to this probability in the following analysis.

\subsection{Control $\left| \hat \Theta_{ij}^{(k)} - \tilde \Theta_{ij}^{(k)}\right|$}\label{th3.1}
Consider the upper bound of entry-wise error term $\left| \hat \Theta_{ij}^{(k)} - \tilde \Theta_{ij}^{(k)}\right|$ for every $(j,k) \in \mathcal A \times [K]$ and $i \in \hat S_j^{(k)} \cup\{j\} = \hat S_{j+}^{(k)}$.
Under all assumptions in Theorem \ref{th: asymp}, we conclude that $ \hat S_j^{(k)}  =  S_j^{(k)} $ simultaneously holds for every $(j,k) \in \mathcal A \times [K]$.
Therefore, in following we use the set $S_{j+}^{(k)}$ instead of $ \hat S_{j+}^{(k)}$.

\subsubsection{Control $\left| \hat \Theta_{j,j}^{(k)} - \tilde \Theta_{j,j}^{(k)}\right|$}\label{th3.1.1}
By the decomposition
\begin{align*}
	\left( \hat \Theta_{jj}^{(k)}\right)^{-1} 
	=& \frac1n \left\| X_j^{(k)} - Z_{\sj}^{(k)} \left( \tilde{\bb}_{\sj}^{(k)} + \delta_{\sj}^{(k)} \right)\right\|_2^2 \\
	= & \frac1n \left\| X_j^{(k)} - X_{S_j^{(k)}}^{(k)}\tilde {\boldsymbol{\alpha}}_{S_j^{(k)}}^{(k)} \right\|_2^2 + \frac1n \left\|Z_{S_j^{(k)}}^{(k)} \delta_{S_j^{(k)}}^{(k)} \right\|_2^2\\
	= & \left( \tilde \Theta_{jj}^{(k)}\right)^{-1} + \frac1n \left\|Z_{S_j^{(k)}}^{(k)} \delta_{S_j^{(k)}}^{(k)} \right\|_2^2 ,
\end{align*}
we have
\begin{equation}\label{eq: jj hat tilde}
	\begin{aligned}
		\left| \hat \Theta_{j,j}^{(k)} - \tilde \Theta_{j,j}^{(k)}\right|
		=& \frac{ \left\|Z_{S_j^{(k)}}^{(k)} \delta_{S_j^{(k)}}^{(k)} \right\|_2^2 \left( \hat \Theta_{jj}^{(k)}\right)^2}
		{n -  \left\|Z_{S_j^{(k)}}^{(k)} \delta_{S_j^{(k)}}^{(k)} \right\|_2^2  \hat \Theta_{jj}^{(k)} }\\
		\le& \frac{8/ \underline C^2}{nN- 4/\underline C }
		\lesssim \frac1{nN} ,
	\end{aligned}
\end{equation}
where the first inequality follows from the EJRIP condition, \eqref{eq: delta}, and $\hat \Theta_{jj}^{(k)} \le 2/ \underline C$ (the result in Appendix \ref{Th1.3.1}).
It also yields that, with a sufficiently large sample size, we have 
\begin{equation}\label{eq: tildejj}
	c_1 \le \tilde \Theta_{j,j}^{(k)} \le c_2 
\end{equation}
where $c_2 > c_1$ are two positive constants.

\subsubsection{Control $\left\| \hat \Theta_{S_j^{(k)},j}^{(k)} - \tilde \Theta_{S_j^{(k)},j}^{(k)}\right\|_2$}\label{th3.1.2}
We have
\begin{equation}\label{eq: Sj hat tilde}
	\begin{aligned}
		&\left\| \hat \Theta_{S_j^{(k)},j}^{(k)} - \tilde \Theta_{S_j^{(k)},j}^{(k)}\right\|_2\\
		\le& \tilde \Theta_{j,j}^{(k)} \left\| \tilde {\boldsymbol{\alpha}}_{S_j^{(k)}}^{(k)} - \hat {\boldsymbol{\alpha}}_{S_j^{(k)}}^{(k)}\right\|_2 + \left| \hat \Theta_{j,j}^{(k)} - \tilde \Theta_{j,j}^{(k)}\right| \cdot \left\| \hat {\boldsymbol{\alpha}}_{S_j^{(k)}}^{(k)}\right\|_2\\
		\overset{(i)}\lesssim & \sqrt{C_0 K} \left\| \left(\hat \Gamma_{S_j^{(k)},S_j^{(k)}}^{(k)}\right)^{-1/2} \right\|_2 \cdot \| \delta_{\sj}\|_2\\
		&+ \sqrt{C_0 K} \left\| \left(\hat \Gamma_{S_j^{(k)},S_j^{(k)}}^{(k)}\right)^{-1/2} \right\|_2 \cdot \left\|\hat \bb_{S_j^{(k)}}^{(k)} \right\|_2 \cdot \frac1{nN} \\
		\overset{(ii)}\lesssim&  \frac{\sqrt K}N + \frac{\sqrt{K}}{nN} \left( \| \hat \bb_{\sj} - \bb_{\sj} \|_2 + \| \bb_{\sj}^{(k)} \|_2 \right) \\
		\lesssim& \frac{\sqrt K}N ,
	\end{aligned}
\end{equation}
where inequality (i) follows from \eqref{eq: jj hat tilde} and \eqref{eq: tildejj}, inequality (ii) follows from Lemma 1, \eqref{eq: delta}, and the last inequality follows from \eqref{eq: sharp error}, \eqref{eq: bjk}, and the sample size assumption.

Therefore, by \eqref{eq: jj hat tilde} and \eqref{eq: Sj hat tilde}, we establish that
\begin{equation}\label{eq: .j hat tilde}
	\max_{(j,k) \in \mathcal A \times [K]} \max_{i \in S_{j+}^{(k)} } \left| \hat \Theta_{ij}^{(k)} - \tilde \Theta_{ij}^{(k)} \right| 
	\le \frac{c_3 \sqrt K}{N}, 
\end{equation}
where $c_3>0$ is a constant. 

\subsection{Asymptotic Normality}\label{th3.2}
Following the proof strategy of Theorem 5, Lemma 3 and Theorem 3 in \cite{shu2024nodewise}, under the assumption $\sqrt n \succ \max_{j \in [p]} s_j \log p$, we obtain that the oracle nodewise estimator $\tilde \Theta_{ij}^{(k)}$ is asymptotically normal as $n \to \infty$, i.e.,
\begin{equation}\label{eq: ora normal}
	\sup_{w\in\mathbb R}
	\left| \mathbf P\left( \frac{\sqrt n \left(\tilde \Theta^{(k)}_{ij} - \Theta^{(k)}_{ij} \right)}{\sigma_{i, S_j}^{(k)}} \le w\right) -\Phi(w) \right|
	=o(1) 
\end{equation}
holds for every $(j,k)\in[p]\times[K]$ and $i\in S_{j+}^{(k)}$, where \begin{equation*}
	\begin{aligned}
		&\left(\sigma_{i, S_j}^{(k)} \right)^2 \\
		:=& \operatorname{Var}\left( \left(\Sigma^{(k)}_{S_{j+}^{(k)}, S_{j+}^{(k)}}\right)^{-1}_{i,\cdot} 
		x_{S_{j+}^{(k)}}^{(k)} \left(x_{S_{j+}^{(k)}}^{(k)}\right)^\top
		\left(\Sigma^{(k)}_{S_{j+}^{(k)}, S_{j+}^{(k)}}\right)^{-1}_{\cdot, j}
		\right).
	\end{aligned}
\end{equation*}

Define $\hat r_{ij}^{(k)} := \frac{\sqrt n (\hat \Theta^{(k)}_{ij} - \tilde \Theta^{(k)}_{ij})}{\sigma_{i, S_j}^{(k)}}$. Then, for every $(j,k) \in \mathcal A \times [K]$, $i \in  S_{j+}^{(k)}$, and $w \in \mathbb R$, we have 
{\small
	\begin{equation}\label{eq: asy1}
		\begin{aligned}
			& \left| \mathbf P\left( \frac{\sqrt n \left(\hat\Theta^{(k)}_{ij} - \Theta^{(k)}_{ij} \right)}{\sigma_{i, S_j}^{(k)}}
			\le w\right) -\Phi(w) \right|\\
			\le&  \left| \mathbf P\left( \frac{\sqrt n \left(\tilde \Theta^{(k)}_{ij} - \Theta^{(k)}_{ij} \right)}{\sigma_{i, S_j}^{(k)}}
			\le w - \hat r_{ij}^{(k)}\right) - \mathbf P\left( \xi\le w - \hat r_{ij}^{(k)}\right) \right| \\
			&+  \left| \mathbf P\left( \xi\le w - \hat r_{ij}^{(k)}\right)- \Phi(w) \right|\\
			\le& \mathbf P\left( |\xi -w| \le  |\hat r_{ij}^{(k)}| \right) + o(1),
		\end{aligned}
	\end{equation}
}

\noindent where $\xi$ denotes an independent $N(0,1)$ random variable, and the last inequality follows from \eqref{eq: ora normal}. 
Furthermore, by \eqref{eq: .j hat tilde} we have 
\begin{equation}\label{eq: asy2}
	\begin{aligned}
		& \mathbf P\left( |\xi -w| \le | \hat r_{ij}^{(k)}|\right) \\
		\le& \mathbf P\left( \left\{|\xi -w| \le | \hat r_{ij}^{(k)}| \right\} \bigcap \left\{  |\hat r_{ij}^{(k)}| \le c_4 \frac1{\sqrt N} \right\} \right)  \\
		&+  \mathbf P\left(  |\hat r_{ij}^{(k)}| > c_4 \frac1{\sqrt N} \right) \\
		\lesssim &  \frac1{\sqrt N} + e^{-C_{2,\gamma} \min_{j \in \mathcal A}\left( \frac{\log p}{s_{0,j}} + \log(s_jK) \right)}.
	\end{aligned}
\end{equation}
Combining \eqref{eq: asy1} and \eqref{eq: asy2}, as $N, K \to \infty$, for every $(j,k) \in \mathcal A \times [K]$ and every $i \in S_{j+}^{(k)}$, we establish that 
$$
\sup_{w\in\mathbb R}
\left| \mathbf P\left( \frac{\sqrt n \left(\hat \Theta^{(k)}_{ij} - \Theta^{(k)}_{ij} \right)}{\sigma_{i, S_j}^{(k)}} \le w\right) -\Phi(w) \right|
=o(1),
$$
which completes the proof of Theorem \ref{th: asymp}.

\section{Technical lemmas} 
\begin{lemma}[Empirical covariance matrices]\label{le: eigen}
	For every $k \in [K]$, assume $X^{(k)} \in \mathbb R^{n \times p}$ is the sample matrix with each row independently following $N(\mathbf 0_p, \Sigma^{(k)})$, and $\hat \Sigma^{(k)} := \frac1n \left(X^{(k)} \right)^\top X^{(k)} \in \mathbb R^{p\times p}$.
	Assume Assumptions 1-3 hold and the sample size of each data set satisfies $n \ge C_1 \max_{j \in [p]}(s_j\log p + s_j s_{0,j} \log K)$.
	With probability $1- C_2 e^{-C_3(\log K + \log p )}$, for every $j \in [p], k \in [K]$, we simultaneously have 
	\begin{equation}\label{eq: opnorm}
		\max_{S \subset [p]: |S|\le C_4 s_j } \left\| \hat \Sigma_{SS}^{(k)} - \Sigma_{SS}^{(k)} \right\|_2 \le C_5 \sqrt{\frac{\log K + s_j\log p}{n}},
	\end{equation}
	\begin{equation}\label{eq: abs} 
		\left| \hat \Sigma_{jj}^{(k)} - \Sigma_{jj}^{(k)} \right| \le C_6 \sqrt{\frac{\log K + \log p}{n}},
	\end{equation}
	and 
	\begin{equation}\label{eq: -1/2}
		\left| \left(\hat \Sigma_{jj}^{(k)}\right)^{-1/2} - \left(\Sigma_{jj}^{(k)}\right)^{-1/2} \right| \le C_7 \sqrt{\frac{\log K + \log p}{n}},
	\end{equation}
	where $C_1, \ldots, C_7$ are positive constants and are independent from $n,p,K,s,\text{ and }s_0$.
\end{lemma}

\begin{proof}[Proof of Lemma \ref{le: eigen}]
	For any set $S$ satisfying $S \subset [p],|S|=C_4 s_j$, by Remark 5.40 in \cite{vershynin2010introduction}, for every $t_j>0$, we have
	\begin{equation}\label{eq: sg design}
		\mathbf P \left( \left\|\frac1n \left(X^{(k)}_S \right)^\top X_S^{(k)} - \Sigma_{SS}^{(k)} \right\|_2 > \max(\delta_j, \delta_j^2)  \right) \le 2e^{-ct_j^2},
	\end{equation}
	where $\delta_j = C\sqrt{\frac {s_j} n } +\frac{t_j}{\sqrt n}$ and $C,c$ are two fixed positive constants.
	
	Then, by taking $t_j =\sqrt{\frac{(C_3+1)\log K + (C_3+C_4+1)s_j \log p }{c} }$ and assuming $n \ge (C \sqrt s_j + t_j)^2$, we get $\delta_j > \delta_j^2$, which yields that
	\begin{equation}\label{eq: maxop}
		\begin{aligned}
			&\mathbf P \left(  \max_{(j,k) \in [p] \times [K]}\max_{S \subset [p]: |S \le C_4 s_j } \left\| \hat \Sigma_{SS}^{(k)} - \Sigma_{SS}^{(k)} \right\|_2 >  \delta_j \right)\\
			\le& 2K \sum_{j \in [p]} \binom p{C_4s_j} \exp\left(-ct_j^2 \right)\\
			\le& 2 \exp\left( -C_3 (\log K + \log p) \right),
		\end{aligned}
	\end{equation}
	where the last inequality follows from $ \binom p{C_4s} < p^{C_4s}$.
	Therefore, we complete the proof of \eqref{eq: opnorm}.
	
	To prove \eqref{eq: abs}, we note that $\hat \Sigma_{jj}^{(k)}= \frac1n \|X_j^{(k)}\|_2^2 \sim \frac{\Sigma_{jj}^{(k)}}n \chi^2(n)$, and by Lemma \ref{le: chi} we have 
	$$
	\mathbf P \left( \left | \hat \Sigma_{jj}^{(k)} - \Sigma_{jj}^{(k)} \right |  >  C \Sigma_{jj}^{(k)} \max\left(\frac tn, \sqrt{\frac tn} \right)  \right) \le 2e^{-t}.
	$$
	We set $t = (C_3+1)\log(pK)$, ensuring $t/n \le \sqrt{t/n}$. By Assumption 2, we learn that $\Sigma_{jj}^{(k)} \in \left[\underline C, \overline C \right]$, which gives
	\begin{equation}\label{eq: maxabs}
		\begin{aligned}
			&\mathbf P \left(   \max_{k \in [K]}\max_{j \in [p]} \left| \hat \Sigma_{jj}^{(k)} - \Sigma_{jj}^{(k)} \right| >  C' \sqrt{ \frac{(C_3+1)\log(pK) }{n}}\right)\\
			\le& 2 pK \cdot \exp\left(- t \right) 
			~=~ 2 \exp\left( -C_3 (\log K + \log p) \right),
		\end{aligned}
	\end{equation}
	confirming inequality \eqref{eq: abs}. 
	
	Additionally, if \eqref{eq: abs} holds and $n$ is sufficiently large, it follows that $\hat \Sigma_{jj}^{(k)} \ge \underline C/{2}$. We obtain
	{\small
		\begin{equation*}
			\begin{aligned}
				&\left| \left(\hat \Sigma_{jj}^{(k)}\right)^{-1/2} - \left(\Sigma_{jj}^{(k)}\right)^{-1/2} \right|
				= \frac{\left| \hat \Sigma_{jj}^{(k)} - \Sigma_{jj}^{(k)} \right|}{\sqrt{\hat \Sigma_{jj}^{(k)} \Sigma_{jj}^{(k)}}\left( \sqrt{\hat \Sigma_{jj}^{(k)} } + \sqrt{ \Sigma_{jj}^{(k)} } \right)} \\
				\le& \frac{2(\sqrt2-1)C' \sqrt{C_3+1}}{{\underline C}^{3/2}} \sqrt{\frac{\log(pK)}n},
			\end{aligned}
		\end{equation*}
	}
	
	\noindent thus we prove inequality \eqref{eq: -1/2} and complete the proof of Lemma \ref{le: eigen}.
\end{proof}

\begin{lemma}[Random error control]\label{le: control}
	For a fixed $j \in [p]$, conditioned on that $Z_{\setminus j}$ satisfies the EJRIP$(C_1s_j, C_2s_{0,j},\delta)$ condition, then with a probability greater than $1- e^{-\frac{C_1' C_2'}{2}s_j s_{0,j} \Delta(s_{0,j} )}$, we have 
	{\small
		\begin{equation*}
			\sup_{S_j \in \mathcal S( C_1' s_j, C_2' s_{0,j})}  
			\left\| \left( \frac{1}{ N} Z_{\setminus j}^\top \epsilon_j \right)_{S_j} \right\|_2^2 \le
			\frac{12\overline C C_1' C_2'}{ N} s_j s_{0,j} \Delta(s_{0,j} ) 
		\end{equation*}
	}
	
	\noindent where $\Delta(s_{0,j} ) := \log K + (1/s_{0,j}) \log p$, and $C_0 \le 1$, $C_1', C_2' \ge 1$ are three constants.
\end{lemma}
\begin{proof}[Proof of Lemma \ref{le: control}]
	By Theorem 2.1 in \cite{hsu12}, for a fixed $S_j \in \mathcal S( C_1' s_j, C_2' s_{0,j})$, a fixed $Z_{\setminus j}$, and every $t >0$, we get
	{\small
		\begin{equation}\label{eq: sg Xi}    \begin{aligned}
				\mathbf P \Bigg\{ &
				\left\| \left( Z_{\setminus j}^\top \epsilon_j \right)_{S_j} \right\|_2^2 \ge 
				\sigma_j^2  tr\left( \left(Z_{\setminus j}^\top Z_{\setminus j}\right)_{S_jS_j}\right) \\
				&+ 2\sigma_j^2 \left\|\left(Z_{\setminus j}^\top Z_{\setminus j}\right)_{S_jS_j} \right\|_F \sqrt t + 2\sigma_j^2 \left\|\left(Z_{\setminus j}^\top Z_{\setminus j}\right)_{S_jS_j} \right\|_2 t  \Bigg\} \le e^{-t},
		\end{aligned}   \end{equation}
	}
	where $\sigma_j^2 : = \max _{k \in [K]} \left(\sigma_j^{(k)} \right)^2 = \max _{k \in [K]} \left(\Theta_{jj}^{(k)} \right)^{-1} \le \overline C$.
	If $Z_{\setminus j}$ satisfies EJRIP condition, we conclude that
	{\small
		\begin{equation*}
			\begin{aligned}
				& tr\left( \left(Z_{\setminus j}^\top Z_{\setminus j}\right)_{S_jS_j}\right) \le   C_1' C_2' s_j s_{0,j} N, \\
				&\left\|\left(Z_{\setminus j}^\top Z_{\setminus j}\right)_{S_jS_j} \right\|_F^2 \le  4C_1' C_2' s_j s_{0,j} N^2,
				\quad \left\|\left(Z_{\setminus j}^\top Z_{\setminus j}\right)_{S_jS_j} \right\|_{2} \le 2N.
			\end{aligned}
		\end{equation*}
	}
	Define the event $\mathcal E_j = \{ Z_{\setminus j} \text{ satisfies EJRIP}(C_1s_j , C_2s_{0,j},\delta) \text{ condition}\}$.
	By taking $ t = \frac32C_1' C_2'  \left(s_j s_{0,j}\log K + s_j \log p \right)$, $C_f := 12\overline C C_1' C_2'$, and using a probability union bound, we get
	{\footnotesize
		\begin{equation*}
			\begin{aligned}
				&\mathbf P\left( \sup_{S_j \in \mathcal S( C_1' s_j, C_2' s_{0,j})}  
				\left\| \left( \frac{1}{ N} Z_{\setminus j}^\top \epsilon_j \right)_{S_j} \right\|_2^2 \ge
				\frac{C_f}{ N} s_j s_{0,j} \Delta(s_{0,j} ) \Big| \mathcal E_j\right)\\ 
				\le & \sum_{S_j \in \mathcal S( C_1' s_j, C_2' s_{0,j})} \mathbf P\left( 
				\left\| \left( \frac{1}{ N} Z_{\setminus j}^\top \epsilon_j \right)_{S_j} \right\|_2^2 \ge
				\frac{C_f}{ N} s_j s_{0,j} \Delta(s_{0,j} ) \Big| \mathcal E_j\right)\\ 
				\le & \binom{p-1}{C_1' s_j} \binom{C_1' s_j K}{ C_1' C_2' s_j s_{0,j}} \exp\left( - \frac32C_1' C_2'  \left(s_j s_{0,j}\log K + s_j \log p \right)\right)\\
				\le & \exp\left(-\frac{C_1' C_2'}{2}s_j s_{0,j} \Delta(s_{0,j} )  \right) ,
			\end{aligned}
		\end{equation*}
	}
	which completes the proof of Lemma \ref{le: control}.
\end{proof}

\begin{lemma}[Chi-squared inequalities]\label{le: chi}
	Let \(x_0, x_1 ,\ldots,x_q\) be \(q+1\) independent central \(\chi^2\) random variables with \(k\) degrees of freedom. 
	Then, for every \(t>0\), there exists an absolute constant \(C>0\) such that
	\begin{equation}\label{eq: chi1}
		\mathbf{P}\left( |x_0-k| \geq C(\sqrt{k t} + t) \right) \leq Ce^{-t},
	\end{equation}
	\begin{equation}\label{eq: chi2}
		\mathbf{P}\left( \frac1{k^2} \sum_{i=1}^q (x_i-k)^2 > 2\left( \frac{q+\sqrt{qt}+t}k + \frac{t^2}{k^2}\right)\right) \leq e^{-t}.
	\end{equation}
	Inequalities \eqref{eq: chi1}, \eqref{eq: chi2} are proved in Lemma 8.1 in \cite{birge2001alternative} and Theorem 3.1 in \cite{bong2023tight} respectively.
\end{lemma}

\section{Extension of the estimation theory to sub-Gaussian designs}
\label{app: sub Gaussian}
Here, we discuss the extension of the joint precision-matrix estimation theory to sub-Gaussian designs.
Consider replacing the normality assumption (Assumption \ref{assume: gauss}) in the main text with the following assumption:
\begin{assumption}[Sub-Gaussianity]\label{assume: sub gauss}
	For each $k \in [K]$, assume $X^{(k)} \in\mathbb{R}^{n_k\times p}$ is row-wise independent and sub-Gaussian: each row $X_{i, \cdot}^{(k)} \overset{d}{=} \zeta_{i}^{(k)}  \left( \Sigma^{(k)} \right)^{1/2}$, where $\zeta_{i}^{(k)} = (\zeta_{i1}^{(k)}, \cdots, \zeta_{ip}^{(k)} ) \in \mathbb{R}^{1 \times p}$ and each $\zeta_{ij}^{(k)}$ is an i.i.d. centered sub-Gaussian random variable such that $\mathbf{E} \left\{ (\zeta_{i}^{(k)})^\top \zeta_{i}^{(k)} \right\} = I_p$, and $\mathbf{E}\left( e^{\lambda \zeta_{ij}^{(k)}} \right) \le e^{\lambda^2 \sigma_X^2/2}$, for every $i\in [n_k]$, $j \in [p]$, $k \in [K]$, and $\lambda \in \mathbb{R}$, where $\sigma_X^2\ge 1$ (as an absolute constant) is the sub-Gaussian parameter of the design.
\end{assumption}

To ensure that the theoretical results established in Section \ref{sec: theoretical} remain valid, we first verify that the nodewise regression error vector, namely $\epsilon_j^{(k)} \in \mathbb{R}^{n_k \times 1}$ in \eqref{eq: node in app}, is element-wise sub-Gaussian for every $k \in [K]$ and $j \in [p]$.
Without loss of generality, we only need to consider the regression error of the first node for any observation $i \in [n_k]$ in any graph $k\in[K]$, denoted as $\epsilon_{i1}^{(k)} =  X_{i1}^{(k)} - X_{i,\setminus1}^{(k)} \boldsymbol{\alpha}_{\setminus1}^{(k)} \in \mathbb R$. 
Here, $X_{i,\setminus1}^{(k)} \in \mathbb R^{1 \times (p-1)}$ represents the remaining subvector of the $i$-th observation row in the $k$-th graph after excluding the first covariate.
Define $\eta_1^{(k)} := \left(1,  -\boldsymbol{\alpha}_{\setminus1}^{(k)\top}\right)^\top \in \mathbb R^{p\times 1}$, we then have
$$
\begin{aligned}
	\mathbf E\left( e^{\lambda \epsilon_{i1}^{(k)} } \right)
	=& \mathbf E\left( e^{\lambda X_{i, \cdot}^{(k)} \eta_1^{(k)}  } \right)
	=\mathbf E \exp\left(  \lambda \zeta_{i}^{(k)}  \left( \Sigma^{(k)} \right)^{1/2} \eta_1^{(k)}   \right)\\
	\le& \exp \left( \frac{\lambda^2\sigma_X^2}2 \left\| \Sigma^{(k)}\right\|_2 \cdot \left\|\eta_1^{(k)} \right\|_2^2 \right),
\end{aligned}
$$
where the last inequality follows from the sub-Gaussianity assumption above. Furthermore, based on the definition of $\boldsymbol{\alpha}_{\setminus1}^{(k)}$ in \eqref{eq: alpha in app}, it follows that
$$
\begin{aligned}
	\left\|\eta_1^{(k)} \right\|_2^2 =& 1+  \left(\Theta_{1, 1}^{(k)}\right)^{-2}  \cdot \left\|\Theta_{\setminus 1, 1}^{(k)}\right\|_2^2\\
	\le& 1+ \Lambda_{\min}^{-2}(\Theta^{(k)})\cdot  \left\|\Theta_{\cdot, 1}^{(k)}\right\|_2^2\\
	\le& 1+ \Lambda_{\max}^{2}(\Sigma^{(k)})\cdot  \left\|\Theta^{(k)}\right\|_2^2\\
	=& 1+ \Lambda_{\max}^{2}(\Sigma^{(k)})\cdot   \Lambda_{\min}^{-2}(\Sigma^{(k)})
	\le 1+ (\overline C/\underline C)^2.
\end{aligned}
$$
Therefore, $\epsilon_{i1}^{(k)}$ follows a sub-Gaussian distribution with its sub-Gaussian parameter $\sigma_j^2 \le \sigma_X^2\left( \overline C + \overline C^3 \underline C^{-2}\right)$ \cite{wainwright2019high}. 
Consequently, constant adjustments are needed in our existing analytical framework.

Finally, the core concentration inequalities used in our lemmas (i.e., \eqref{eq: sg design} from Remark 5.40 in \citep{vershynin2010introduction}, and \eqref{eq: sg Xi} from Theorem 2.1 in \cite{hsu12}) remain valid under sub-Gaussian designs. 
Similarly, the relevant quadratic form inequalities (i.e., \eqref{eq: chi2} from Theorem 3.1 in \cite{bong2023tight}, and \eqref{eq: chi1} from Lemma 8.1 in \cite{birge2001alternative}) also hold for sub-exponential variables. 
Thus, the sub-Gaussian Assumption \ref{assume: sub gauss} does not affect the convergence rate analysis, and hence the theoretical results in Section \ref{sec: theoretical} remain valid under sub-Gaussian designs.

\section{Additional simulation results}\label{app: simu}

We provide further numerical experiments in this appendix, covering algorithm performance under additional measurements, robustness to heavy-tailed distributions, and simulations across more network structures.

\subsection{Additional results for the simulation in Section \ref{sec 4.1}}
We present supplementary results for the simulation studies in Section \ref{sec 4.1}.
All parameters are the same as in Section \ref{sec 4.1}. 
In terms of estimation accuracy, we additionally evaluate the maximum column-wise $L_1$-norm (Max L1 Norm), defined as $\max_{j \in [p]} \sum_{k \in [K]} \frac{n_k}{N} \left\| \hat{\Theta}^{(k)}_{\cdot j} - \Theta^{(k)}_{\cdot j} \right\|_1$. 
Regarding support set recovery, we incorporate Precision, Recall, F1 score, and Hamming Loss, calculated as follows:
\begin{equation*}
	\begin{aligned} 
		\text{Precision} =& \frac{TP}{TP+ FP},\\
		\text{Recall} =& \frac{TP}{TP+ FN},\\
		\text{F1 score} =& \frac{2}{(\text{Precision})^{-1}+ (\text{Recall})^{-1}},\\
		\text{Hamming loss} =& FP+FN,
	\end{aligned}
\end{equation*}
where TP denotes true positives, FP denotes false positives, and FN denotes false negatives.
We compute these four quantities at two levels: (i) edge-level selection accuracy across all $K$ graphs (reported as Precision-Edge, Recall-Edge, F1-Edge, Hamming-Edge), and (ii) neighbor-level selection accuracy using the union $S_j = \bigcup_{k\in [K]} S_j^{(k)}$ for each node $j \in [p]$ (reported as Precision-Ngbr, Recall-Ngbr, F1-Ngbr, Hamming-Ngbr). 
All reported values are averages over 100 Monte Carlo replications. Higher Precision/Recall/F1 and lower Max L1 Norm/Hamming loss indicate better performance.

Table \ref{tab: simuadd} presents the supplementary results.
All Precision/Recall/F1 values have been multiplied by 100, and standard errors are shown in parentheses.
Consistent with Table 1 in Section \ref{sec 4.1}, across all examined values of $\rho$, our MIGHT method attains competitive estimation accuracy compared to competing methods. 
In terms of the edge-level and the neighbor-level support recovery, the MIGHT estimator preserves relatively high Precision and attains the best overall performance on composite metrics such as F1 score and Hamming loss. 

\begin{sidewaystable}
	\centering
	\caption{Additional measurements across varying $\rho$.}
	\label{tab: simuadd}
	\resizebox{\textheight}{!}{\begin{tabular}{@{}lccccccccc@{}} 
\hline
			& \multicolumn{1}{c}{Estimation} & \multicolumn{4}{c}{Edge-wise Support Recovery} & \multicolumn{4}{c}{Neighbor-wise Support Recovery} \\
			\cmidrule(lr){2-2} \cmidrule(lr){3-6} \cmidrule(lr){7-10}
			Methods & Max L1 Norm &Precision-Edge & Recall-Edge & F1-Edge & Hamming-Edge & Precision-Ngbr& Recall-Ngbr & F1-Ngbr& Hamming-Ngbr  \\
\hline
			\multicolumn{2}{c}{ $\rho = 0.2$} & & & \\
			MIGHT      & {\bf 8.023 (0.112) }   & {\bf98.266 (0.068)} & 44.479 (0.443) & {\bf61.202 (0.424)} & {\bf4412.000 (68.402)}  & 98.797 (0.096) & 83.739 (0.547)  & {\bf90.619 (0.327)} & {\bf170.067 (6.506)}    \\
			GGL        & 10.588 (0.124)   & 49.007 (0.542) & 53.213 (0.496) & 50.926 (0.290) & 8029.333 (112.060)  & 25.086 (1.176) & 97.219 (0.980)  & 39.294 (1.058) & 3040.933 (109.350) \\
			JEM        & 9.527 (0.147)    & 94.766 (0.146) & 36.176 (0.656) & 52.268 (0.703) & 5155.933 (83.513)   & {\bf99.855 (0.040)} & 52.077 (0.800)  & 68.347 (0.700) & 470.800 (10.502)   \\
			FJEM       & 310.986 (12.031)  
			& 8.130 (0.069)  & {\bf95.529 (0.241)} & 14.983 (0.117) & 84806.800 (299.661)  
			& 9.888 (0.086)  & {\bf100.000 (0.000) } & 17.993 (0.143) & 8921.067 (8.536)   \\
			Sep Glasso & 13.169 (0.130)   & 48.674 (0.304) & 24.573 (0.335) & 32.618 (0.314) & 7932.200 (75.785)   & 33.437 (0.311) & 78.050 (0.533)  & 46.768 (0.283) & 1740.800 (23.391)  \\
			Sep Node   & 11.835 (0.138)   & 96.191 (0.262) & 5.954 (0.089)  & 11.210 (0.159) & 7377.600 (69.475)   & 95.383 (0.305) & 37.898 (0.459)  & 54.203 (0.492) & 626.533 (9.288)    \\ \hline
			\multicolumn{2}{c}{ $\rho = 0.5$} & & & \\
			MIGHT      & {\bf 4.645 (0.089) }   & 96.415 (0.088)  & 54.565 (0.634) & {\bf69.628 (0.521)} & {\bf 2341.600 (44.850)}   & 98.776 (0.099) & 79.159 (0.588)  & {\bf87.850 (0.357)} & {\bf215.800 (6.790)}    \\
			GGL        & 6.917 (0.123)    & 52.231 (0.653) & 67.697 (0.917) & 58.802 (0.523) & 4675.267 (95.499)   & 45.193 (1.611) & 87.357 (1.218)  & 58.645 (1.412) & 1301.933 (112.747) \\
			JEM        & 6.142 (0.096)    & 88.103 (0.230) & 42.651 (0.991) & 57.292 (0.903) & 3112.267 (60.960)   & {\bf99.752 (0.074)} & 53.247 (0.943)  & 69.293 (0.805) & 462.733 (11.435)   \\
			FJEM       &  305.408 (16.413) 
			& 5.180 (0.042)  & {\bf96.528 (0.231)} & 9.831 (0.075)  & 87189.600 (392.454) 
			& 9.949 (0.078)  & {\bf100.000 (0.000) } & 18.095 (0.128) & 8915.067 (7.679)   \\
			Sep Glasso & 8.716 (0.119)    & 49.300 (0.323) & 37.201 (0.622) & 42.309 (0.404) & 4982.133 (51.580)   & 35.931 (0.345) & 83.046 (0.634)  & 50.094 (0.311) & 1633.600 (27.844)  \\
			Sep Node   & 7.340 (0.139)    & {\bf97.050 (0.225)} & 10.790 (0.217) & 19.401 (0.351) & 4409.733 (41.578)   & 96.462 (0.278) & 42.081 (0.688)  & 58.508 (0.672) & 586.533 (10.120)   \\ \hline
			\multicolumn{2}{c}{ $\rho = 0.8$} & & & \\
			MIGHT      & {\bf1.767 (0.032)}    & 94.462 (0.184) & 70.161 (0.496) & {\bf80.488 (0.335)} & {\bf673.800 (12.248)}    & 97.949 (0.148) & 73.731 (0.460)  & {\bf84.106 (0.294)} & {\bf245.933 (4.865) }   \\
			GGL        & 2.786 (0.041)    & 37.528 (0.486) & 87.525 (0.429) & 52.441 (0.432) & 3163.067 (68.096)   & 45.296 (0.398) & 87.855 (0.493)  & 59.707 (0.298) & 1048.133 (14.790)  \\
			JEM        & 2.582 (0.037)    & 83.763 (0.404) & 60.347 (0.774) & 70.029 (0.449) & 1020.933 (14.343)   & {\bf99.659 (0.069)} & 59.727 (0.772)  & 74.607 (0.594) & 357.533 (7.459)    \\
			FJEM       & 513.932 (36.266)  
			& 2.084 (0.019)  & {\bf98.942 (0.172) } & 4.082 (0.037)  & 92248.667 (469.228) 
			& 8.918 (0.062)  & {\bf100.000 (0.000)} & 16.374 (0.105) & 9017.133 (6.150)   \\
			Sep Glasso & 3.708 (0.044)    & 46.726 (0.306) & 68.719 (0.602) & 55.566 (0.243) & 2178.333 (25.182)   & 37.463 (0.284) & 87.771 (0.464)  & 52.471 (0.243) & 1405.933 (19.041)  \\
			Sep Node   & 2.581 (0.039)    & {\bf94.732 (0.312)} & 31.719 (0.382) & 47.491 (0.442) & 1388.600 (13.522)   & 93.738 (0.380) & 53.858 (0.575)  & 68.365 (0.511) & 439.267 (6.891)    \\
\hline
\end{tabular}
}
\end{sidewaystable}

\subsection{Robustness analysis}
To evaluate the robustness of the proposed method, we have conducted additional simulations based on $t$-distributions with varying degrees of freedom ($df$).
Following the Erd\H{o}s--R\'{e}nyi graph setting in Section \ref{sec 4.1} (with $\rho=1/2$), we set $n_k=100$, $p=100$, and $K=10$. 
The $\ell$-th observation in the $k$-th graph is generated as 
$$
x_\ell^{(k)} = (\Theta^{(k)})^{-1/2} \xi_\ell^{(k)} \in \mathbb R^p,
$$
where $\xi_\ell^{(k)}$ is independent $p$-dimensional $t$-distributed vector for every $\ell \in [n_k]$ and $k \in [K]$. For a fair comparison, the $t$-distributions are rescaled to have unit variance across all $df$ settings.

Table \ref{tab: robust} summarizes the results based on 50 independent replications. 
Our regression-based MIGHT method demonstrates competitive robustness, consistently outperforming competing methods across all considered degrees of freedom ($df$). 
As the $df$ increases, the estimation error of MIGHT decreases, and the accuracy of structure recognition improves.
Notably, at $df=9$, the performance of MIGHT is comparable to its Gaussian results (reported in Table 1 of the main manuscript), which aligns with the fact that a $t$-distribution with large degrees of freedom approximates a normal distribution. 

\begin{table}[!t]
\centering
\begin{threeparttable} 
	\caption{Performance across $t$-distributions with different degrees of freedom.}
	\label{tab: robust}
	\footnotesize
	\setlength{\tabcolsep}{1.2pt} 
	\renewcommand{\arraystretch}{1.0}
	\begin{tabular}{@{}lcccc@{}} 
		\toprule
		& \multicolumn{2}{c}{Estimation Accuracy} & \multicolumn{2}{c}{Structure Recognition} \\
		\cmidrule(lr){2-3} \cmidrule(lr){4-5}
		Methods & Frobenius & Max L2  &
		MCC-Edge & MCC-Ngbr \\
		\midrule
		\multicolumn{2}{c}{ $t(df=3)$} & & & \\
		MIGHT      & {\bf15.842} (0.115)  & {\bf4.750} (0.141)      & {\bf63.445} (0.801) & {\bf79.015} (0.751) \\
		GGL        & 17.755 (0.288)  & 5.174 (0.191)      & 49.855 (1.082) & 47.849 (2.546) \\
		JEM        & 18.376 (0.196)  & 5.480 (0.145)      & 57.531 (1.103) & 69.818 (0.992) \\
		FJEM       & 195.859 (4.510) & 2188.3 (165.9)  
		& 3.594 (0.034) & 0.000 (0.000)\\
		Sep Glasso & 24.990 (0.177)  & 9.190 (0.160)      & 36.184 (0.513) & 43.724 (0.552) \\
		Sep Node   & 18.290 (0.106)  & 6.303 (0.139)      & 30.808 (0.491) & 59.841 (0.774)  \\ \hline
		\multicolumn{2}{c}{ $t(df=6)$} & & & \\
		MIGHT      & {\bf14.044} (0.129)  & {\bf3.476} (0.101)      & {\bf67.734} (0.552) & {\bf81.948} (0.486) \\
		GGL        & 17.545 (0.283)  & 5.015 (0.153)      & 53.964 (0.784) & 48.335 (2.696) \\
		JEM        & 18.776 (0.159)  & 5.492 (0.099)      & 60.319 (0.788) & 71.669 (0.776) \\
		FJEM       & 201.072 (3.059) & 2946.7 (157.9)  
		& 3.904 (0.031) & 0.000 (0.000)\\
		Sep Glasso & 25.575 (0.175)  & 9.571 (0.149)      & 39.406 (0.477) & 46.535 (0.413) \\
		Sep Node   & 18.287 (0.113)  & 6.213 (0.110)      & 31.342 (0.395) & 61.606 (0.616) \\ \hline
		\multicolumn{2}{c}{ $t(df=9)$} & & & \\
		MIGHT      & {\bf13.736} (0.118)  & {\bf3.437} (0.098)      & {\bf68.953} (0.574) & {\bf82.908} (0.551) \\
		GGL        & 17.850 (0.183)  & 5.162 (0.109)      & 56.745 (0.692) & 55.294 (2.104) \\
		JEM        & 18.646 (0.127)  & 5.582 (0.098)      & 60.184 (0.807) & 71.484 (0.775) \\
		FJEM       & 181.285 (2.963) & 2095.3 (168.6)  
		& 3.897 (0.072) & 0.000 (0.000)\\
		Sep Glasso & 25.272 (0.143)  & 9.377 (0.141)      & 40.391 (0.451) & 47.679 (0.407) \\
		Sep Node   & 18.075 (0.096)  & 6.198 (0.128)      & 31.998 (0.312) & 62.367 (0.567) \\ 
		\bottomrule
	\end{tabular}
\end{threeparttable} 
\end{table}

\subsection{Simulations under more network structures}
To provide a more comprehensive assessment, we further investigate the performance of MIGHT under diverse network structures and a broader range of parameter combinations, under a larger $p=200$.
Specifically, we consider the following four joint graphical structures:

\begin{enumerate}
\item \textbf{Erd\H{o}s--R\'{e}nyi:} Similar to the setting in Section \ref{sec: simu}, we first generate a base graph $G$ where each pair of nodes is connected with a probability of $0.05$. Each individual graph $G_k$ is then generated by randomly removing a proportion $\rho = 1/2$ of the edges from $G$.

\item \textbf{Band:} Motivated by \cite{Guo11joint, Jankova2017honest}, we construct a base graph $G$ that only connects node pairs $(i,j)$ satisfying $|i-j| \le 2$. 
Each individual graph is formed by randomly removing a proportion $\rho = 1/2$ of the edges from $G$.

\item \textbf{Block Hub:} 
We first generate three base graphs. In each base graph, the $p$ variables are partitioned into 40 blocks of equal size, and one node is randomly selected as the hub within each block and connected to all other nodes in that block.
To form an individual graph, we partition the adjacency matrix into a non-bottom-right part and a bottom-right part. Inspired by \cite{Ma16joint}, we independently select one from the three base graphs for the non-bottom-right part, and again independently select one from the three base graphs for the bottom-right part; the corresponding submatrices are taken directly from the chosen base graphs.
This mixing procedure ensures that individual graphs share structural similarity (all parts come from the same pool of base graphs) while also exhibiting heterogeneity (different combinations of base graphs).

\item \textbf{Block Scale-free:}
We generate three base graphs, each partitioned into 4 diagonal blocks. Within each block, a scale-free network is generated using the R package \texttt{igraph}. 
Individual graphs are then constructed in the same way as in \textbf{Block-Hub}: the non-bottom-right part and the bottom-right part are each obtained by randomly selecting one from the three base graphs (independently for the two parts) and taking the corresponding submatrix.
\end{enumerate}

\begin{table*}[!h]
\centering 
\captionsetup{justification=centering}
\caption{Estimation Accuracy with Different structures and $(n_k, K)$.} 
\label{tab: estimation R3}
\renewcommand{\arraystretch}{1.0}  
\setlength{\tabcolsep}{4pt}    
\resizebox{\linewidth}{!}{\begin{tabular}{@{}lcccccccccc@{}}  
		\toprule%
		& \multicolumn{2}{c}{MIGHT} & \multicolumn{2}{c}{GGL} & \multicolumn{2}{c}{JEM} & \multicolumn{2}{c}{Sep Glasso} & \multicolumn{2}{c}{Sep Node} \\
		\cmidrule(lr){2-3} \cmidrule(lr){4-5} \cmidrule(lr){6-7} \cmidrule(lr){8-9} 
		\cmidrule(lr){10-11}  
		Structure & Frobenius  & Max L2
		& Frobenius  & Max L2
		& Frobenius  & Max L2
		& Frobenius  & Max L2
		& Frobenius  & Max L2\\
		\midrule
		\multicolumn{2}{r}{$(n_k, K)= (100,5)$ }& & & & & & & & &  \\
		Erd\H{o}s--R\'{e}nyi  & {\bf21.215}      & {\bf4.625}        & 28.300      & 7.240       & 31.054    & 8.622     & 37.868           & 11.149            & 26.839         & 7.775           \\
		Band & {\bf9.721}       & {\bf1.376}       & 13.430     & 1.562      & 15.123    & 2.005     & 20.866           & 3.149             & 13.993         & 1.925           \\
		Block Hub  & {\bf6.663}       & {\bf0.814}        & 8.665     & 0.897      & 9.038     & 0.958     & 18.099           & 3.124             & 10.095         & 1.412           \\
		Block Scale-free &  {\bf13.405} & {\bf2.966} & 15.906 &  4.479 & 17.372 & 5.182 &  22.764 &  6.609&   16.164 & 6.049 \\
		\multicolumn{2}{r}{$(n_k, K)= (100,10)$}& & & & & & & & &  \\
		Erd\H{o}s--R\'{e}nyi   & {\bf20.328}      & {\bf3.764}        & 27.385    & 6.280       & 28.245    & 6.429     & 37.885           & 10.780             & 26.696         & 7.162           \\
		Band & {\bf8.839}       & {\bf0.911}        & 12.651    & 1.262      & 12.064    & 1.270      & 20.862           & 2.923             & 14.019         & 1.614           \\
		Block Hub  & {\bf6.327}       & 0.551        & 8.132     & 0.733      & 7.084     & {\bf0.512}     & 18.093           & 3.048             & 10.130          & 1.380            \\
		Block Scale-free & {\bf12.511} & {\bf2.676} &   15.401 & 3.947 & 15.622 &  3.698 & 22.655 &  6.454&  16.114 & 5.772\\
		\multicolumn{2}{r}{$(n_k, K)= (200,10)$}& & & & & & & & &  \\
		Erd\H{o}s--R\'{e}nyi  & {\bf12.616}      & {\bf1.489}        & 24.393    & 4.863      & 28.743    & 6.715     & 32.250            & 7.658             & 22.963         & 4.564           \\
		Band & {\bf5.457}       & {\bf0.354}        & 12.514    & 1.198      & 11.988    & 1.240      & 17.314           & 2.051             & 10.043         & 0.973           \\
		Block Hub  & {\bf3.883}       & {\bf0.209}       & 7.847     & 0.687      & 6.761     & 0.452     & 14.760            & 2.095             & 5.740        & 0.548  \\ 
		Block Scale-free & {\bf7.792} & {\bf1.070} &    14.247 & 3.255     & 15.461 &  3.605      & 18.643 & 4.172      &     12.512  & 2.810 \\
		\bottomrule
	\end{tabular}
} 
\end{table*}

\begin{table*}[!h]
\centering 
\captionsetup{justification=centering}
\caption{Structure Recognition with Different structures and $(n_k, K)$.} 
\label{tab: structure R3}
\renewcommand{\arraystretch}{1.0}  
\setlength{\tabcolsep}{4pt}    
\resizebox{\linewidth}{!}{\begin{tabular}{@{}lcccccccccc@{}}  
		\toprule%
		& \multicolumn{2}{c}{MIGHT} & \multicolumn{2}{c}{GGL} & \multicolumn{2}{c}{JEM} & \multicolumn{2}{c}{Sep Glasso} & \multicolumn{2}{c}{Sep Node} \\
		\cmidrule(lr){2-3} \cmidrule(lr){4-5} \cmidrule(lr){6-7} \cmidrule(lr){8-9} 
		\cmidrule(lr){10-11}  
		Structure & MCC-Edge & MCC-Ngbr
		& MCC-Edge & MCC-Ngbr
		& MCC-Edge & MCC-Ngbr
		& MCC-Edge & MCC-Ngbr
		& MCC-Edge & MCC-Ngbr\\
		\midrule
		\multicolumn{2}{r}{$(n_k, K)= (100,5)$ }& & & & & & & & &  \\
		Erd\H{o}s--R\'{e}nyi   & {\bf62.552} & {\bf68.291} & 51.083 & 50.695 & 36.869 & 41.746 & 33.830  & 42.945 & 26.741 & 40.984 \\
		Band & {\bf85.431} & {\bf88.704} & 59.645 & 54.270  & 70.287 & 74.935 & 53.933 & 61.860  & 51.823 & 71.900   \\
		Block Hub  & {\bf93.445} & 93.805 & 31.858 & 28.152 & 89.414 & {\bf94.036} & 67.568 & 58.242 & 69.159 & 83.256 \\
		Block Scale-free & {\bf64.164} & {\bf64.580} & 44.141 &  43.108 &  47.846 & 46.867 & 43.053 &  44.546 & 36.330 & 45.293 \\
		\multicolumn{2}{r}{$(n_k, K)= (100,10)$}& & & & & & & & &  \\
		Erd\H{o}s--R\'{e}nyi  & {\bf65.614} & {\bf78.369} & 57.405 & 58.896 & 52.135 & 61.437 & 33.197 & 47.590  & 26.955 & 54.501 \\
		Band & {\bf89.116} & {\bf97.217} & 55.045 & 44.422 & 84.814 & 94.279 & 54.315 & 58.752 & 51.640  & 84.148 \\
		Block Hub  & {\bf94.728} & {\bf96.186} & 26.832 & 20.070  & 88.701 & 94.844 & 67.148 & 47.831 & 69.096 & 89.928 \\
		Block Scale-free & {\bf68.551} & {\bf72.473}  &    47.449 & 45.727   &     57.088  & 59.036  &   43.825 & 45.099  &   36.599 & 53.111\\
		\multicolumn{2}{r}{$(n_k, K)= (200,10)$}& & & & & & & & &  \\
		Erd\H{o}s--R\'{e}nyi & {\bf88.454} & {\bf96.112} & 62.098 & 54.196 & 51.069 & 59.132 & 55.089 & 50.994 & 46.088 & 81.232 \\
		Band & {\bf96.617} & {\bf98.626} & 80.507 & 73.411 & 87.974 & 94.498 & 67.052 & 48.887 & 73.799 & 97.142 \\
		Block Hub  & {\bf98.652} & 98.381 & 46.705 & 37.015 & 91.357 & 98.420  & 65.966 & 40.000     & 89.881 & {\bf98.476} \\ 
		Block Scale-free & {\bf89.015} & {\bf91.022}  &   60.460 & 55.992    &  59.164&  58.780   &    61.303&  54.708    &   59.734 & 78.185  \\  
		\bottomrule
	\end{tabular}
} 
\end{table*}

In the Erd\H{o}s--R\'{e}nyi and Band settings, for each individual graph $G_k$, edge weights are sampled uniformly from $[-1, -0.5] \cup [0.5, 1]$ to form the adjacency matrix $\Omega^{(k)}$. The corresponding precision matrix $\Theta^{(k)}$ is constructed as:
$$\Theta^{(k)} = \Omega^{(k)} + \left(0.1 + |\Lambda_{\min}(\Omega^{(k)})| \right)\cdot \mathbf I_p, 
\quad \text{ for every }k \in [K],
$$ 
with $p=200$.
In the Block-Hub and Block-Scale-free settings, only the three base graphs are generated using the above scheme.
Each individual graph is then constructed by combining sub-blocks selected from these base graphs. 
Thus, for a given neighboring pair $i-j$, the edge strength \(\Theta_{ij}^{(k)}\) across the \(K\) graphs can take only the three values specified by the base graphs, thereby incorporating a degree of value similarity.

We use the same competing methods and evaluation metrics as in Section \ref{sec: simu}.  
The fast and scalable joint estimator (FJEM) exhibited relatively poor numerical stability in the preceding experiments, with large estimation errors and weak structure-recovery performance.
In the \(p=200\) simulations, FJEM continued to show numerical instability and extremely large estimation errors; therefore, we do not include it in the resulting tables.
Tables \ref{tab: estimation R3} and \ref{tab: structure R3} summarize the results for $(n_k, K) = (100, 5), (100, 10), \text{ and } (200, 10)$, based on 50 independent replications. The results demonstrate that the proposed MIGHT method outperforms competing methods across nearly all considered settings.
Notably, increasing the number of related individual graphs ($K$) facilitates more accurate identification of shared structures (Table \ref{tab: structure R3}) and improves joint estimation accuracy (Table \ref{tab: estimation R3}).
Likewise, increasing the sample size per graph ($n_k$) leads to better estimation accuracy and support recovery.

\end{document}